\documentclass[aps,prb,superscriptaddress,twocolumn]{revtex4-1}

\usepackage{graphics}
\usepackage{graphicx} 
\usepackage{dcolumn}
\usepackage{bm}
\usepackage{epstopdf}
\usepackage[usenames, dvipsnames]{color}
\usepackage{amsmath}

\begin{document}


\title{Magnetic exchange interactions in the van der Waals layered antiferromagnet MnPSe$_3$}



\author{S.~Calder}
\email{caldersa@ornl.gov}
\affiliation{Neutron Scattering Division, Oak Ridge National Laboratory, Oak Ridge, Tennessee 37831, USA.}

\author{A.~V.~Haglund}
\affiliation{Department of Materials Science and Engineering, University of Tennessee, Knoxville, TN 37996.}

\author{A.~I.~Kolesnikov}
\affiliation{Neutron Scattering Division, Oak Ridge National Laboratory, Oak Ridge, Tennessee 37831, USA.}

\author{D.~Mandrus}
\affiliation{Department of Materials Science and Engineering, University of Tennessee, Knoxville, TN 37996.}
\affiliation{Materials Science and Technology Division, Oak Ridge National Laboratory, Oak Ridge, TN 37831.}

 
\date{\today}

\begin{abstract}	
Two-dimensional van der Waals compounds with magnetic ions on a honeycomb lattice are hosts to a variety of exotic behavior. The magnetic interactions in one such compound, MnPSe$_3$, are investigated with elastic and inelastic neutron scattering. Magnetic excitations are observed in the magnetically ordered regime and persist to temperatures well above the ordering temperature, $\rm T_N$ = 74 K, consistent with low dimensional magnetic interactions. The inelastic neutron scattering results allow a model spin Hamiltonian to be presented that includes dominant intralayer interactions of $J_{1ab}$=0.45 meV, $J_{2ab}$=0.03 meV, $J_{3ab}$=0.19 meV, and appreciable interlayer interactions of $J_c$=0.031(5) meV. No evidence for anisotropy in the form of a spin-gap is observed in the data collected. The measurements on MnPSe$_3$ are contrasted with those on MnPS$_3$ and reveal a large increase in the interlayer exchange interactions in MnPSe$_3$, despite the quasi-2D magnetic behavior. 
\end{abstract}

\maketitle

\section{Introduction}

Materials composed of two-dimensional (2D) layers, either in bulk compounds or reduced to a few or single layers, provide avenues for enhanced quantum phenomena. In this context 2D monolayer graphene, formed from the exfoliation of weakly connected van der Waals (vdW) bonded layers in graphite, has ignited widespread interest due to a variety of novel behavior \cite{NatureGraphene,ZhangQHallGraphene, Yu61}. General 2D material research has extended beyond graphene to the broader class of cleavable vdW materials \cite{doi:10.1021/acsnano.5b05556}. A particular focus has been on materials with magnetic ions on 2D layers with predictions of topological and enhanced quantum behavior in the bulk compounds down to the monolayer. In the 2D atomic layer limit for Heisenberg spins the Mermin-Wagner theorem prohibits magnetic order \cite{PhysRevLett.17.1133}, however, Onsager's theorem for Ising spins supports the stabilization of long range 2D order, which would apply to materials with strong magnetocrystalline anisotropy \cite{PhysRev.65.117}. Therefore, an understanding of the exchange interactions and anisotropy provides indications on the stability and behavior of low dimensional magnetism.

The class of 2D vdW materials has begun expanding and includes Fe$\rm _{3-x}$GeTe$_2$ \cite{FGTdoi:10.1002/ejic.200501020, PhysRevB.93.014411, ISI:000442526400011, ISI:000430389100006, PhysRevLett.122.217203,PhysRevB.99.094423, Wangeaaw8904}, Cr$X$Te$_3$ ($X$=Si,Ge,Sn) \cite{Carteaux_1995, PhysRevB.92.144404, PhysRevLett.123.047203, PhysRevB.100.060402}, CrPS$_4$ \cite{LOUISY197861, PhysRevB.102.024408}, CrCl$_3$ \cite{PhysRevMaterials.1.014001} and VI$_3$ \cite{PhysRevB.100.094402,PhysRevB.99.041402}. Ferromagnetism in atomically thin 2D layers has been shown in CrI$_3$ and CrGeTe$_3$ \cite{doi:10.1021/acs.nanolett.6b03052, ISI:000402823400033, ISI:000402823400032}. One of the initial classes of 2D vdW materials is $M$P$X$$_3$ ($M$=Fe, Mn, Co, Ni and $X$=S, Se)\cite{PhysRevB.46.5425,Rule_2009,Wildes_2012,PhysRevB.76.134402,Wildes_2017,PhysRevLett.121.266801}. These are characterized by 2D hexagonal layers of magnetic ions and host behavior including observation of antiferromagnetism in single layer FePS$_3$ \cite{doi:10.1021/acs.nanolett.6b03052} to ferrotoroidicity in MnPS$_3$ \cite{PhysRevB.82.100408}. 

MnPSe$_3$ belongs to this class of magnetic 2D vdW materials; however, it has undergone limited research, particularly when compared to the sulphur analog MnPS$_3$.  MnPSe$_3$ is a semiconductor with an optical energy gap of 2.27 eV measured from absorption data \cite{Grasso:99}. This is consistent with earlier reports from optical absorption of 2.5 eV \cite{Brec1979doi:10.1021/ic50197a018}. MnPSe$_3$ shows potential for valleytronics, with a study of single layer MnPSe$_3$ on YMnO$_3$  \cite{PhysRevApplied.11.014011}. MnPSe$_3$ has magnetic ions forming the same hexagonal motif as graphene with crystal space group R$\bar{3}$ ($\# 148$) \cite{WIEDENMANN1981}. Antiferromagnetic order of the Mn$^{2+}$ ions ($3d^5$), in the high-spin state S=5/2, has been reported at 74 $\pm$2 K from neutron scattering \cite{WIEDENMANN1981}. The size of the magnetic and crystallographic unit cell are the same due to the propagation vector of {\bf k}=(000). The reduced T$\rm _N$ compared to a Curie-Weiss temperature of -201 K \cite{ISI:A1982NN00100006} supports the low dimensional nature of the magnetism.  Considering the reduction to monolayer theoretically the magnetic ground state of 2D MnPSe$_3$ remains antiferromagnetic with a predicted ordering temperature of 88 K, close to the value of the bulk system \cite{doi:10.1021/ja505097m}. Experimental indications show that the magnetic ordering temperature remains similar from bulk to a few nanometer \cite{ISI:000522634300031}. Considering the magnetic structure in Ref.~\onlinecite{WIEDENMANN1981} for bulk MnPSe$_3$ concludes from neutron diffraction that the spins are confined to the $ab$-plane but cannot rule out a non-zero component along the $c$-axis. A more recent study using powder diffraction also concluded the spins for MnPSe$_3$ are confined to the basal plane \cite{PhysRevMaterials.4.034411}. The moment direction has implications for an understanding of the anisotropic, Ising, XY or 3D Heisenberg nature of the spins, and any competition between single ion anisotropy and dipolar anisotropy. 

Comparisons of MnPSe$_3$ and related MnPS$_3$ have led to intriguing findings which form broader context for this study. Despite the analogous Mn$^{2+}$ 2D lattice and similar magnetic ordering temperatures, T$\rm _N$(MnPSe$_3$)=74 K and T$\rm _N$(MnPS$_3$)=78 K, the magnetism is altered from predominantly isotropic in MnPS$_3$ to indications of unusually large XY anisotropy in MnPSe$_3$ along with a flipping of the spin direction \cite{Jeevanandam_1999}. The magnetic anisotropy in Ref.~\onlinecite{Jeevanandam_1999} of MnPSe$_3$ was proposed as a consequence of the zero-field splitting of the Mn-ion ground state (S=5/2, L=0) from a combination of spin-orbit coupling and the axial crystal field due to a trigonal distortion of the MnS$_6$ octahedra. This octahedral structural distortion, however, is reduced in MnPSe$_3$ compared to MnPS$_3$. Indeed, there is debate in the literature regarding the origin of the anisotropy in MnPS$_3$. A spin-gap at the magnetic Brillouin zone center of 0.5 meV was measured using Inelastic Neutron Scattering (INS) \cite{WildesMnPS3}. This was modeled within a framework of linear spin-wave theory with the introduction of an easy-axis term to account for the anisotropy, rather than a dipolar coupling term. It was noted that the calculated difference between the two routes was negligible for the data. One signature of the dipole-dipole coupling is a splitting of the magnon excitation \cite{PICH199530}. Indications of this was observed with the high resolution available of neutron spin-echo \cite{HICKS2019512}. Consequently, MnPS$_3$ is best described as hosting both single-ion anisotropy and dipolar coupling, with continued research aimed at further insights, for example through Raman studies \cite{VaclavkovaRaman}.

Here we present powder neutron diffraction and INS measurements on MnPSe$_3$ to access the anisotropy and exchange interactions in this bulk vdW layered material. Magnetic excitations, consistent with spin waves, are observed in the magnetically ordered regime that dampen but persist above $\rm T_N$. Analysis of the data provides a model spin Hamiltonian that requires the inclusion of both intralayer interactions, $J_{1ab}$, $J_{2ab}$, $J_{3ab}$ and  interlayer exchange interaction $J_c$. The data does not reveal any observable spin-gap, indicating weak anisotropy. The results reveal that while the exchange interactions in  MnPSe$_3$  are dominated by intralayer interactions, the appreciable interlayer ($J_c$) interactions contrast to that observed for MnPS$_3$, with an alteration of an order of magnitude. 

\section{Experimental Details}

\subsection{Sample synthesis}

Powdered samples of MnPSe$_3$ and MnPS$_3$ were synthesized through annealing of their constituent elements: Mn (99.95$\%$, Alfa Aesar), P (99.999$+\%$, Alfa Aesar), and Se (99.999$\%$, Alfa Aesar) for MnPSe$_3$, or S (99.9999$\%$, Puratronic) for MnPS$_3$. For each sample, under argon atmosphere to avoid oxygen and moisture effects, stoichiometric amounts of the elements were measured out to equal 6 g, and ground in an agate mortar and pestle until reaching a fine texture and uniform color. The fine powder was then loaded into a pellet press die and transferred to ambient atmosphere for pressing. Once compressed, the resulting pellet was sealed in a fused-quartz tube under a constant vacuum of 10-2 Torr, before being placed in a muffle furnace for annealing. To avoid overpressure of the tube due to volatilization of P, S, and Se, the samples were slowly heated to 730$^{\circ}$ C over 3 days, then held at that temperature for 1 week, before allowing to cool to room temperature. The resulting fine powders of bright green MnPS$_3$ and wine red MnPSe$_3$ were verified for correct crystal structure on an Empyrean X-ray diffractometer (Malvern Panalytical).

\subsection{Neutron powder diffraction}
  
Neutron powder diffraction was carried out on the HB-2A powder diffractometer at the High Flux Isotope Reactor (HFIR), Oak Ridge National Laboratory (ORNL) \cite{Garlea2010, doi:10.1063/1.5033906}. A vertically focusing germanium monochromator was used to select a wavelength of  1.12 $\rm \AA$ from the Ge(117) reflection and 1.54 $\rm \AA$ from the Ge(115) reflection. HB-2A has a Debye–-Scherrer geometry and the diffraction pattern was collected by scanning a 120$^{\circ}$ bank of 44 $^3$He detectors in 0.05$^{\circ}$ steps to give 2$\theta$ coverage from 5 to 130$^{\circ}$. Soller collimators of 12$'$ are located before each detector. Measurements with a 1.12 $\rm \AA$ wavelength were collected at 295 K, 100 K, 75 K and 15 K for 4 hours. Measurements with a 1.54 $\rm \AA$ wavelength were collected from 15 to 110 K in 5 K temperature steps for 1 hour.  A 5 gram sample of MnPSe$_3$ was contained in a vanadium sample holder with diameter 0.9 cm. The data was normalized using a vanadium measurement. The wavelengths and detector positions were calibrated with a Si measurement. The neutron data collected was reduced using Mantid \cite{ARNOLD2014156}.  

\subsection{Inelastic neutron scattering}

INS measurements were performed on the time-of-flight direct geometry spectrometer Sequoia \cite{1742-6596-251-1-012058} at the Spallation Neutron Source, ORNL. Data were collected on MnPSe$_3$ and MnPS$_3$ powder samples. The samples were loaded into a 1 cm diameter cylindrical aluminium can and measured utilizing the 3-sample changer in the closed cycle refrigerator (CCR). An identical empty aluminium can was measured and used to subtract the background from the sample environment. Data were collected with incident energies of E$\rm _i$=20 meV and E$\rm _i$=8 meV. Both of these energies were in high resolution mode. For E$\rm _i$=20 meV the Fermi chopper frequency was 240 Hz with a T$_0$ chopper frequency of 60 Hz giving an energy resolution at the elastic line of 0.48 meV. For E$\rm _i$=8 meV the Fermi chopper frequency was 120 Hz with a T$_0$ chopper frequency of 30 Hz  giving an energy resolution at the elastic line of 0.18 meV. The energy resolution was calculated using rez.mcvine.ornl.gov and confirmed by taking cuts of the elastic line in the data. The momentum resolution, Q, was based on fitting the resolution of the Bragg peaks at the elastic line. Measurements were collected for 6 hours at 200, 100, 70 and 15 K. Shorter measurements of 1 hour from 22 K to 95 K in steps of 7$^{\circ}$ were also collected with $\rm E_i$=8 meV. The neutron data collected were reduced using Mantid \cite{ARNOLD2014156}. Data reduction and analysis utilized the DAVE software \cite{DAVE} and modeling was performed with SpinW  \cite{spinW}.

\section{Results and Discussion}

\subsection{Neutron powder diffraction measurements}

\begin{figure}[tb]
	\centering         
	\includegraphics[trim=0cm 8.5cm 0.5cm 0cm,clip=true, width=1.0\columnwidth]{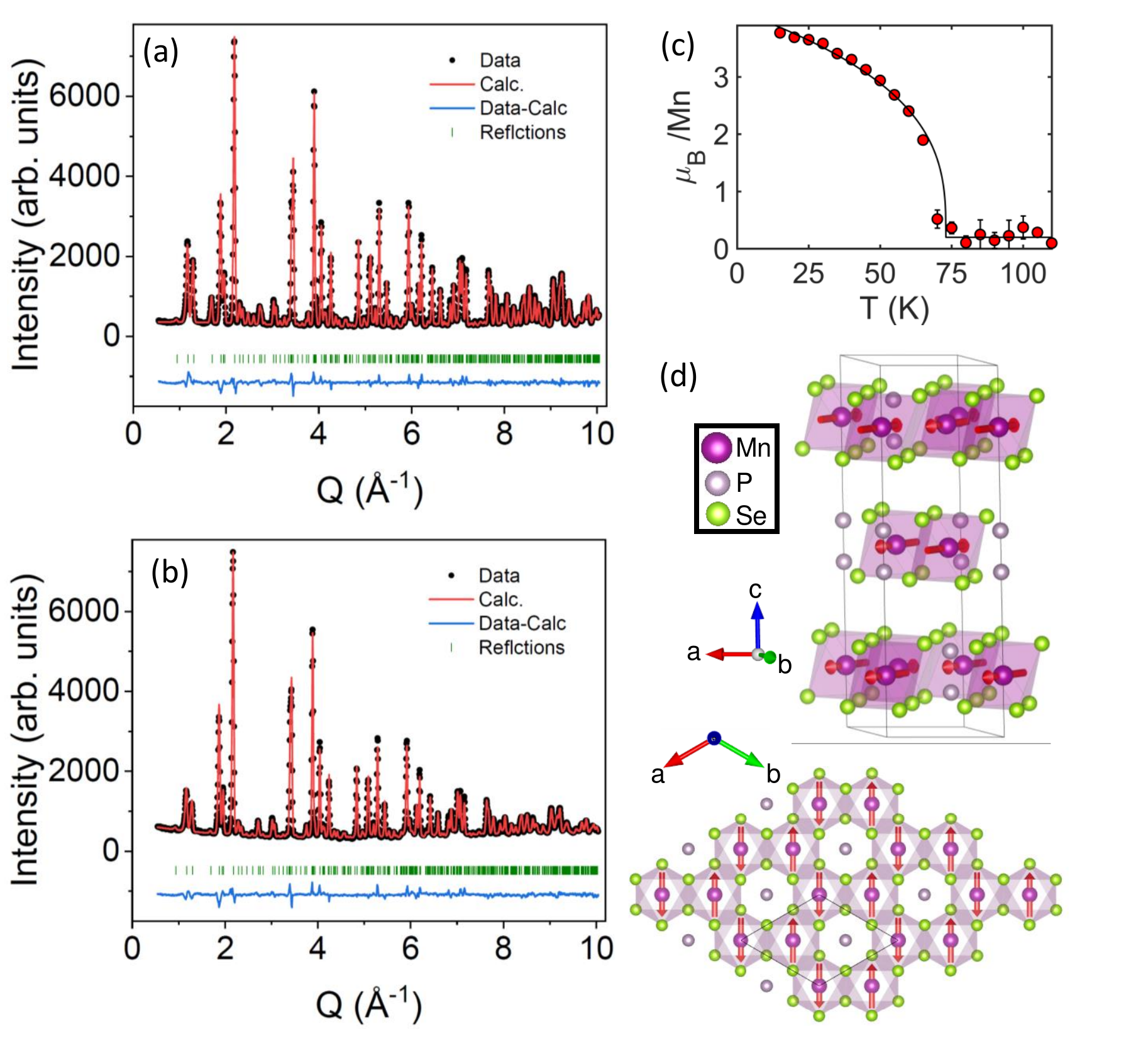}           
	\caption{\label{Mag_Str} Refinements of neutron powder diffraction measurements on MnPSe$_3$ with a wavelength of $\lambda=1.12$ $\rm \AA$ at (a) 15 K and (b) 295 K. (c) Refined Mn$^{2+}$ magnetic moment at 15 K to 110 K in 5K temperature steps through the magnetic ordering transition. The solid line is a fit to a power law. (d) Crystal and magnetic structure of MnPSe$_3$ produced using VESTA \cite{VESTA}. The unit cell is indicated by the solid lines.}
\end{figure}

Neutron powder diffraction measurements were performed on MnPSe$_3$ to probe the crystal and magnetic structure as a function of temperature, as shown in Fig.~\ref{Mag_Str}. To initially investigate the crystal symmetry a short wavelength of 1.12 $\rm \AA$ was used to give a large Q coverage up to 10 $\rm \AA^{-1}$. This allowed detailed measurements of the lattice, atomic positions and thermal parameters with Rietveld refinement, which was performed using the Fullprof software \cite{Fullprof}. The refinements did not require the inclusion of any preferred orientation parameter.  No indication of any structural symmetry lowering from the R$\bar{3}$ ($\# 148$) was found with these measurements, or with the measurements using the longer wavelength of 1.54 $\rm \AA$. MnPSe$_3$ consists of Mn-layers stacked in an $ABC$ stacking sequence separated by 6.65 $\rm \AA$ at 15 K. The Se layers are stacked in an $ABAB$ sequence. The  trigonally distorted MnSe$_6$ octahera have Mn-Se bond distances of 2.757(8)$\rm \AA$ and 2.721(7)$\rm \AA$. 

\begin{table}[htb]
	\begin{tabular}{ccc|rrr}
		IR  &  BV  &  Atom & \multicolumn{3}{c}{BV components}\\
		&      &             &$m_{\|a}$ & $m_{\|b}$ & $m_{\|c}$ \\
		\hline
		$\Gamma_{1}$ & \mbox{\boldmath$\psi_1$} &      1 &      0 &      0 &      1   \\
		&                                       &      2 &      0 &      0 &      1 \\
		$\Gamma_{2}$ & \mbox{\boldmath$\psi_2$} &      1 &      0 &      0 &      1   \\
		&                                       &      2 &      0 &      0 &     -1   \\
		$\Gamma_{3}$ &\mbox{\boldmath$\psi_3$}  &      1 &      1 &      1 &      0  \\
		&                                       &      2 &      1 &      1 &      0   \\
		$\Gamma_{4}$ & \mbox{\boldmath$\psi_4$} &      1 &      1 &      1 &      0  \\
		&                                       &      2 &     -1 &     -1 &      0 \\
		$\Gamma_{5}$ & \mbox{\boldmath$\psi_5$} &      1 &  1 & -1 &      0     \\
		&                                       &      2 &  1 & -1 &      0     \\
		$\Gamma_{6}$ & \mbox{\boldmath$\psi_6$} &      1 &  1 & -1 &      0 \\
		&                                       &      2 & -1 &  1 &      0  \\
	\end{tabular}
	\caption{Basis vectors for the space group R$\bar{3}$:H with 	${\bf k}=(000)$ in Kovalev's notation. The decomposition of the magnetic representation for 	the Mn site $( 0,~ 0,~ .17484)$ is $\Gamma_{Mag}=1\Gamma_{1}^{1}+1\Gamma_{2}^{1}+1\Gamma_{3}^{1}+1\Gamma_{4}^{1}+1\Gamma_{5}^{1}+1\Gamma_{6}^{1}$. The atoms of the nonprimitive basis are defined according to 1: $( 0,~ 0,~ 0.16817)$, 2: $( 0,~ 0,~ 0.83183)$.}
	\label{IRtable}
\end{table}

Turning now to the magnetic structure. As previously reported measurements indicated, the spins are in the $ab$-plane with a propagation vector {\bf k}=(000) \cite{WIEDENMANN1981, PhysRevMaterials.4.034411}. We consider this further based on symmetry arguments using two complimentary approaches: Representational analysis and magnetic space groups. The representational analysis approach was performed using the SARAh software \cite{sarahwills}. This assumes the high temperature paramagnetic space group remains unaltered in the magnetic phase. This is consistent with the refinements of the crystal structure at all temperatures measured, see Fig.~\ref{Mag_Str}(a). Under this case the irreducible representations (IRs) contain magnetic structures with spins either confined to the $ab$-plane or magnetic structures with spins along the $c$-axis, see Table \ref{IRtable}. To allow for a magnetic structure with a component of the spin out of either the basal plane or $c$-axis requires the combination of more than one IR. This goes against Landau theory which states that for a continuous transition only one IR is required. The IR $\Gamma_{4}$ was used to determine the magnetic structure in this approach. Conversely, using the magnetic space group approach we find no maximally allowed magnetic space group compatible with the data. The only maximally allowed magnetic space groups found with the Bilbao Crystallographic Server \cite{Bilbao_Mag} have spins only along the $c$-axis. Therefore we were required to go to a further lower symmetry subgroup to find the magnetic space group of P-1' ($\#2.6$), in BNS notation. Within this magnetic space group the spins are symmetry-allowed to have non-zero and different magnetic components along the $a$-,$b$- and $c$-axis. Results from both the IR and magnetic space group approaches were refined against the data. No distinction was observed and therefore we follow the highest symmetry approach since no breaking of the $a$=$b$ symmetry was measured. The magnetic structure therefore confines the spins to the $ab$-plane, with arbitrary direction within the plane. The refined magnetic moment was determined to be 3.83(3) $\mu\rm _B$$/$Mn$^{2+}$.  We note that the P-1' magnetic space group allows both a definite spin direction within the ab-plane, as well as a component along the c-axis, moreover this space group supports magnetoelectric coupling. Further measurements will be of interest to probe for symmetry breaking within the layer to provide more insight into the magnetic structure.

To follow the evolution of the magnetic ordering, the magnetic and crystallographic scattering were refined from measurements in 5 K steps from 15 to 110 K, shown in Fig.~\ref{Mag_Str}(c). No anomalies of the crystal lattice were observed within this temperature range. The temperature dependence of the ordered Mn$^{2+}$ moment was fit to a power law, shown in Fig.~\ref{Mag_Str}. The onset of long range order is consistent with the reported transition temperature of 74 K. The best fit critical exponent was $2\beta$=0.14(1). This is consistent with 2D ordering, as has been observed in related bulk vdW compounds \cite{PhysRevB.92.224408}.

\begin{figure}[tb]
	\centering         
	\includegraphics[trim=0cm 0cm 0cm 0cm,clip=true, width=1.0\columnwidth]{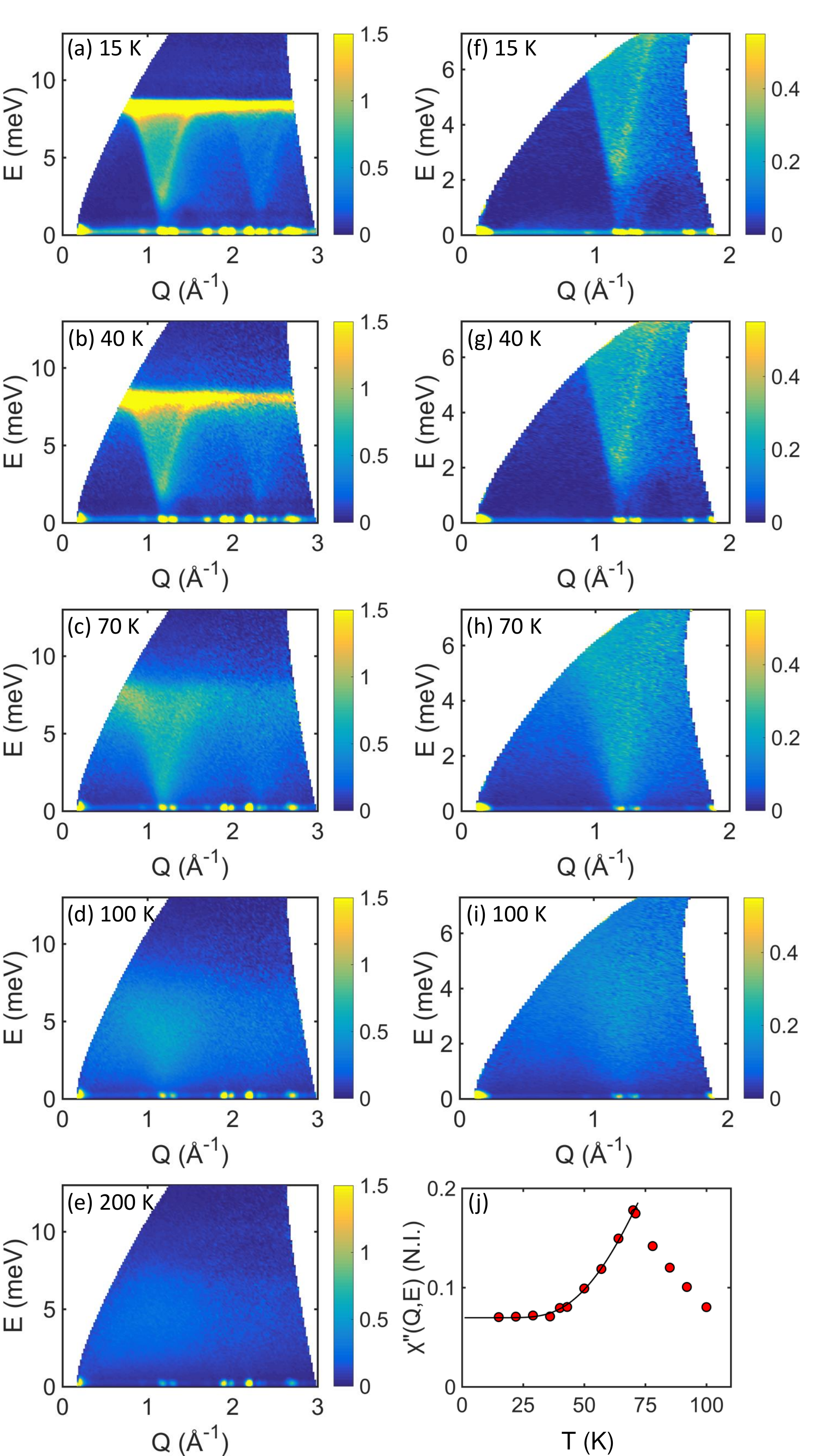}           
	\caption{\label{MnPSe3_INS_data} INS data for MnPSe$_3$. E$\rm _i$=20 meV Bose factor corrected data at (a) 15 K, (b) 40 K, (c) 70 K, (d) 100 K and (e) 200 K. E$\rm _i$=8 meV Bose factor corrected data at (f) 15 K, (g) 40 K, (h) 70 K, (i) 100 K. (j) Bose corrected intensity in region E=0.3-0.8 meV and Q=1.135-1.235 $\rm \AA^{-1}$ to follow the evolution of the low energy scattering around the magnetic Bragg point.}
\end{figure} 

\begin{figure}[tb]
	\centering         
	\includegraphics[trim=0cm 0cm 0cm 0cm,clip=true, width=1.0\columnwidth]{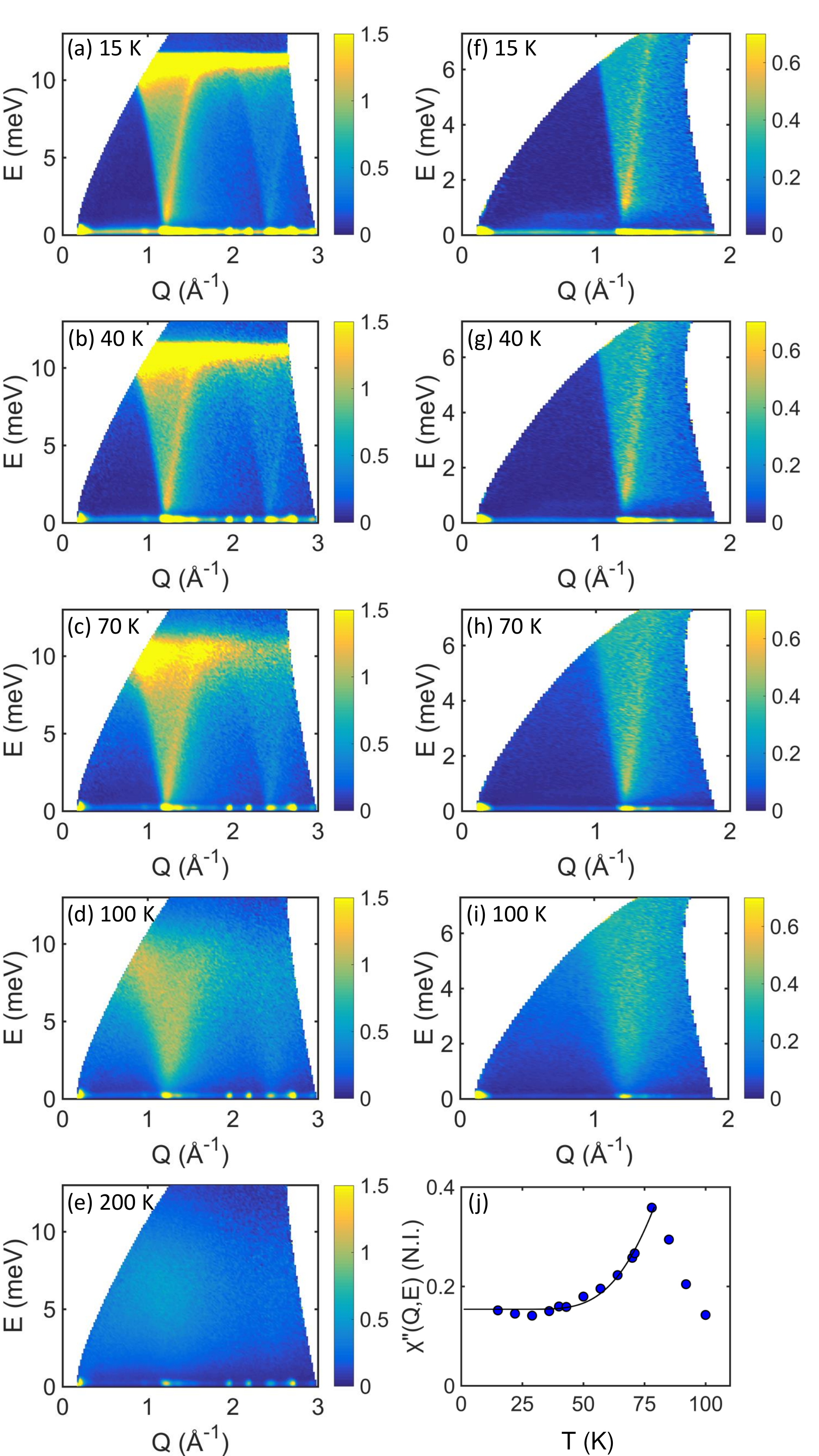}           
	\caption{\label{MnPS3_INS_data} INS data for MnPS$_3$. E$\rm _i$=20 meV at (a) 15 K, (b) 40 K, (c) 70 K, (d) 100 K and (e) 200 K. E$\rm _i$=8 meV at (f) 15 K, (g) 40 K, (h) 70 K, (i) 100 K. (j) Bose corrected intensity in region E=0.3-0.8 meV and Q=1.17-1.27 $\rm \AA^{-1}$ to follow the evolution of the low energy scattering around the magnetic Bragg point.}
\end{figure} 

\section{Inelastic neutron scattering measurements}

To access the magnetic exchange interactions and anisotropy, inelastic neutron scattering measurements were performed on MnPSe$_3$ at various temperatures through the $\rm T \rm _N$ = 74 K magnetic ordering transition, see Fig.~\ref{MnPSe3_INS_data}. Complimentary data, shown in Fig.~\ref{MnPS3_INS_data}, at the same incident energy and temperature conditions were collected for MnPS$_3$, $\rm T _N$ = 78 K, to allow for a direct comparison of the data between MnPSe$_3$ and MnPS$_3$. The MnPS$_3$ scattering within the magnetically ordered phase, as expected, is consistent with the single crystal data in the literature \cite{WildesMnPS3}. The data presented here allows the temperature dependence to be observed.

The data shown in Fig.~\ref{MnPSe3_INS_data} for the inelastic scattering in MnPSe$_3$ have temperature and Q dependence behavior consistent with magnetic scattering. The well defined excitations within the magnetically ordered regime broaden but remain robust at 100 K and even persist to 200 K, the highest temperatures measured. This temperature behavior indicates low dimensional exchange interactions and the presence of short range correlations to high temperatures. Contrasting the INS data for MnPSe$_3$ with MnPS$_3$ shows a reduced energy scale with a maximum energy of 8.5 meV for MnPSe$_3$ compared to 11.5 meV for MnPS$_3$, despite the similar ordering temperatures. Nevertheless, similar temperature behavior is observed for MnPS$_3$ and MnPSe$_3$, with well defined magnetic excitations dampening, but remaining present above the magnetic transition. 

\begin{figure}[tb]
	\centering         
	\includegraphics[trim=0cm 8.5cm 2.5cm 0cm,clip=true, width=0.8\columnwidth]{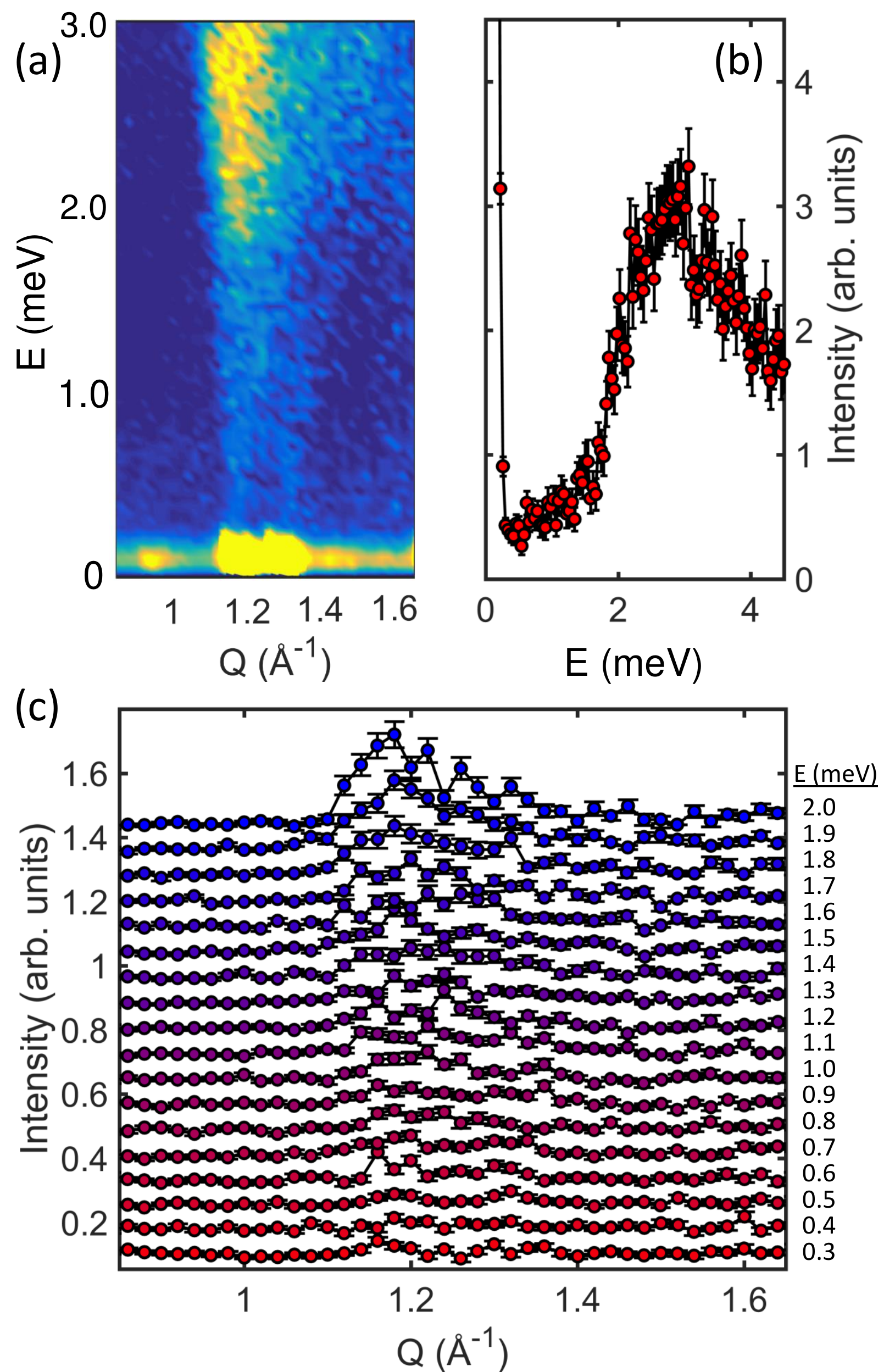}           
	\caption{\label{Fig_MnPSe3_Low_E_1dcuts} (a) Low energy inelastic scattering of MnPSe$_3$ at the magnetic Bragg points with E$\rm _i$=8 meV at 15 K. (b) Inelastic intensity in the Q range 1.12 $\rm \AA^{-1}$ to 1.22 $\rm \AA^{-1}$. (c) Constant energy cuts with an energy step of 0.1 meV. The energies of each cut are indicated. The cuts have been offset by a constant factor.}
\end{figure}

Focusing on low energy and inspecting the data at the magnetic Bragg positions (Q=1.2-1.3 $\rm \AA^{-1}$) reveals another difference in the spectra of MnPSe$_3$ as MnPS$_3$. The MnPS$_3$ inelastic scattering signal extends down to E=0.5 meV, see Fig.~\ref{MnPS3_INS_data}(a) and Fig.~\ref{MnPS3_INS_data}(f) and Ref.~\onlinecite{WildesMnPS3}. For MnPSe$_3$ the low energy magnetic scattering is qualitatively different, with strong scattering cutting off below 2 meV. This can be seen most clearly in the S(Q, $\omega$) scattering in Fig.~\ref{Fig_MnPSe3_Low_E_1dcuts}(a) and  constant momentum Fig.~\ref{Fig_MnPSe3_Low_E_1dcuts}(b) and constant energy Fig.~\ref{Fig_MnPSe3_Low_E_1dcuts}(c) cuts taken at low energy in MnPSe$_3$. While it is tempting to assign this behavior to a spin-gap due to increased anisotropy, as predicted from magnetic susceptibility measurements \cite{Jeevanandam_1999}, closer inspection reveals weak scattering dispersing from the magnetic Bragg points at all energies. Therefore any spin-gap is smaller than the energy resolution. Instead, as we show below when the data are modeled, the magnetic excitation spectra  indicates an alteration of the interlayer interactions between MnPSe$_3$ and MnPS$_3$.

In Fig.~\ref{MnPSe3_INS_data}(h) and Fig.~\ref{MnPS3_INS_data}(h) the Bose corrected intensity in a fixed Q and energy transfer window are plotted to follow the low energy temperature dependence through $\rm T_N$. The  T$<$$\rm T_N$ results are fit to $\chi''(T) \propto(-\Delta/k_BT)$. A pronunced change in the spectral wieght occurs at the expected ordering temperatures consistent with T$\rm _N$(MnPSe$_3$)=74 K and  T$\rm _N$(MnPS$_3$)=78 K.

\subsection{Magnetic Exchange Interactions}

\begin{figure}[bt]
	\centering         
	\includegraphics[trim=0cm 2cm 0cm 0cm,clip=true, width=1.0\columnwidth]{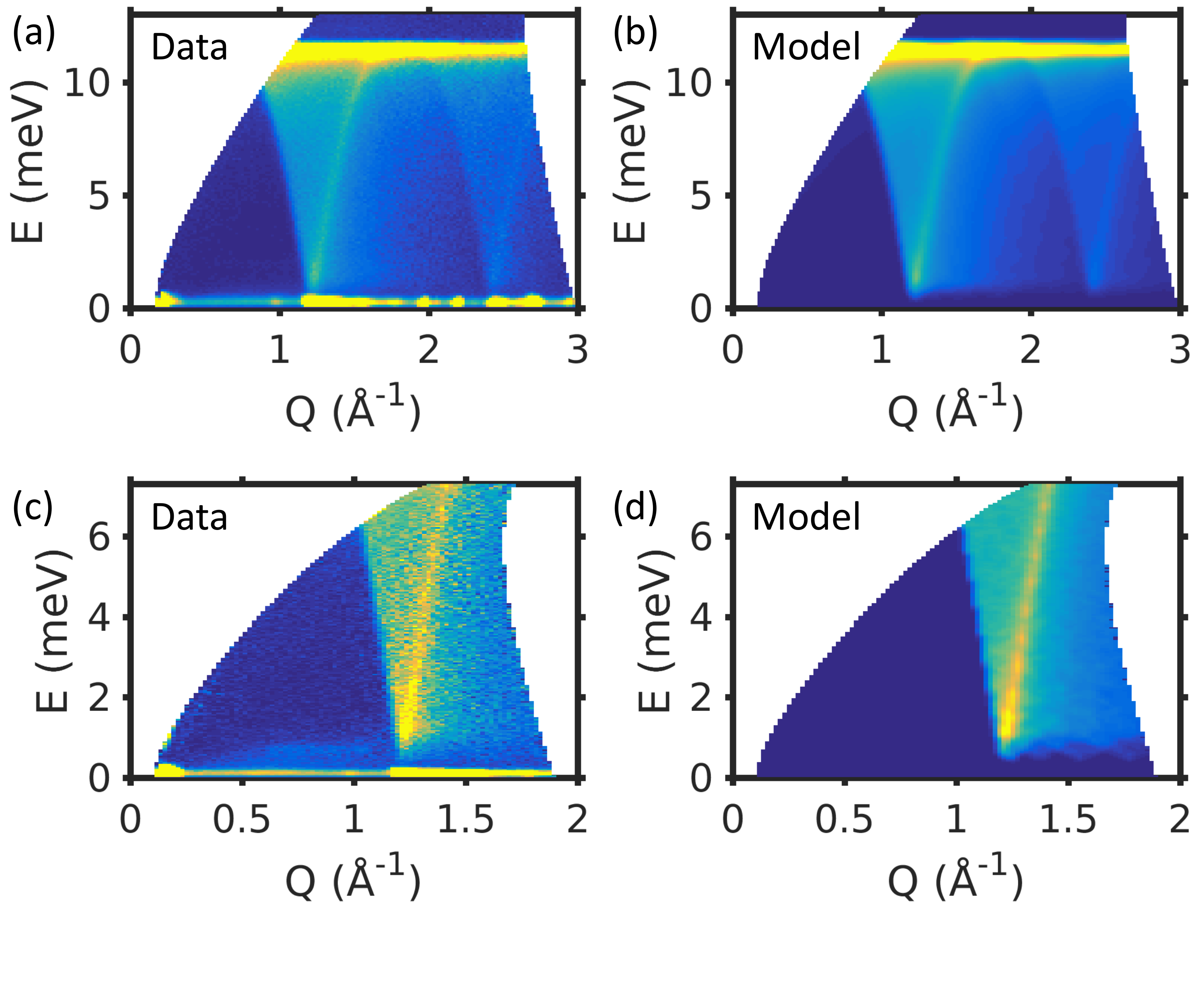}           
	\caption{\label{MnPS3_INS_models} INS data and spin wave model for MnPS$_3$ based on previously determined exchange interactions \cite{WildesMnPS3}. Data collected at 15 K with (a) an $\rm E_i$= 20 meV and (b) corresponding model simulation. (c) Data collected at 15 K with an $\rm E_i$= 8 meV and (d) corresponding model simulation.}
\end{figure} 

\begin{figure*}[tb]
	\centering         
	\includegraphics[trim=0cm 9.5cm 1cm 0cm,clip=true, width=0.9\textwidth]{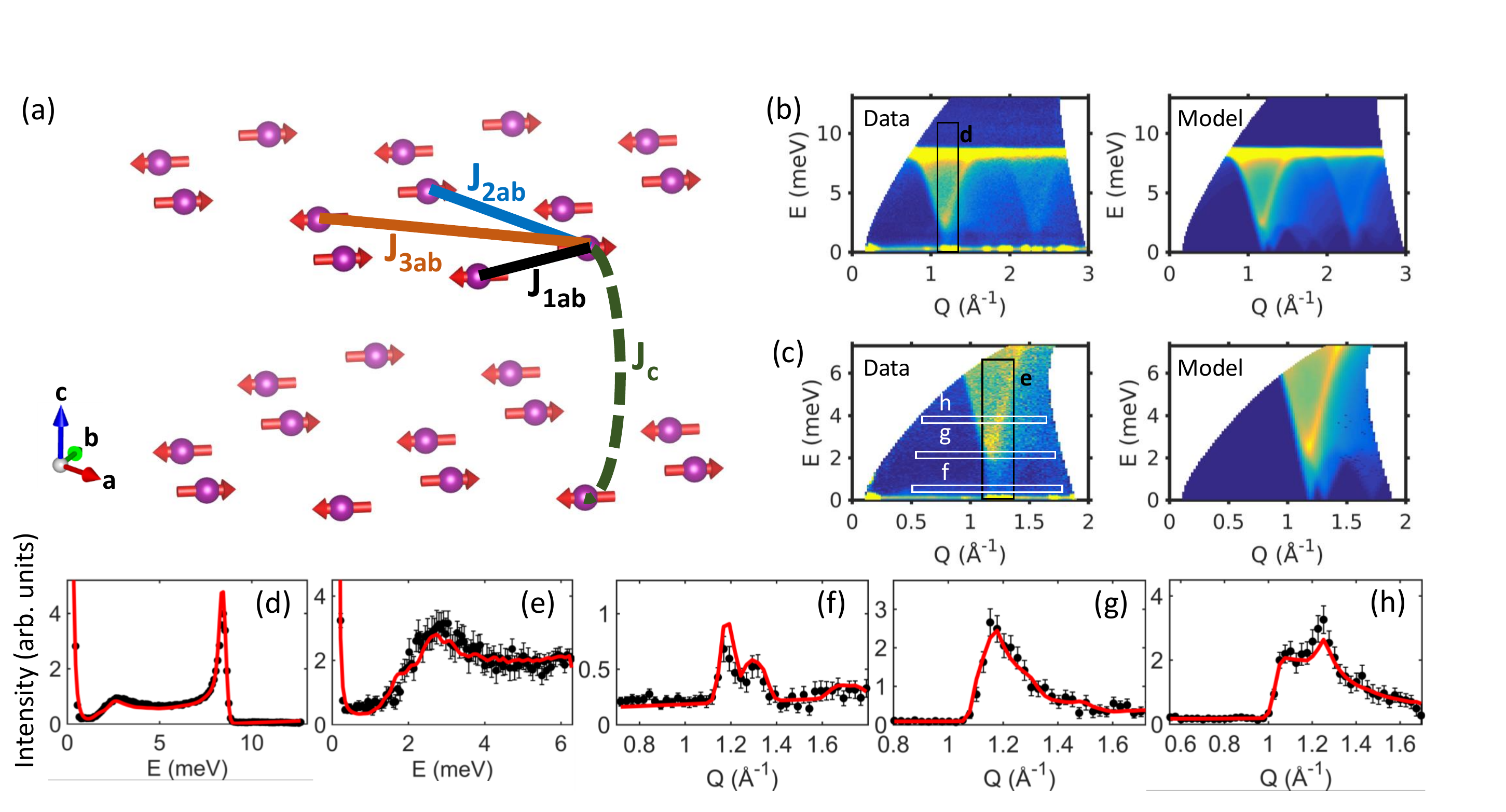}           
	\caption{\label{MnPSe3_INS_models} (a) Exchange interactions between the Mn ions (purple spheres) in MnPSe$_3$ used in the spin Hamiltonian. Inelastic neutron scattering data and model based on the best fit exchange interaction parameters for (b) $\rm E_i$= 20 meV and  (c) $\rm E_i$= 8 meV. All data was collected at 15 K. The boxes highlight the data cuts shown in the rest of the figure. Data (circles) and model fit (line) for (d) cut along energy with $\rm E_i$=20 meV in the range 1.12 $\le$ Q $(\rm \AA^{-1}$) $\le$ 1.22.  (e) Cut along energy with $\rm E_i$=8 meV in the range 1.12 $\le$ Q $(\rm \AA^{-1}$) $\le$ 1.22.  (f) Cut along Q with $\rm E_i$=8 meV in the range 0.5 $\le$ E (meV) $\le$ 0.8. (g) Cut along Q with $\rm E_i$=8 meV in the range 2.0 $\le$ E (meV) $\le$ 2.3. (h) Cut along Q with $\rm E_i$=8 meV in the range 3.7 $\le$ E (meV) $\le$ 4.0.}
\end{figure*}

To gain a quantitative understanding of the magnetic excitations we turn to modeling the data with linear spin wave theory to extract exchange interactions and anisotropy. Inelastic neutron scattering on single crystals is the premier method to achieve this due to the well understood scattering cross section S(Q, $\omega$) and access to four dimensional energy and momentum space with Q=H,K,L. Indeed this was done on MnPS$_3$ and so the fitting is not repeated here for MnPSe$_3$. The reproduced results are shown in Fig.~\ref{MnPS3_INS_models} to confirm the validity of the model for the MnPS$_3$ powder sample. However, we note that for low dimensional vdW compounds the hierarchy of intralayer and interlayer exchange interactions can allow the extraction of detailed Hamiltonians from the S($|$Q$|$, $\omega$) powder data, as has been shown in previous  studies \cite{0953-8984-24-41-416004, PhysRevB.98.134414, PhysRevB.102.024408, INSCoPS3}. We therefore follow this approach to provide a spin Hamiltonian for MnPSe$_3$. Moreover, the high symmetry of MnPSe$_3$, as confirmed above with neutron powder diffraction, simplifies the required exchange parameter further and  makes the development of a model spin Hamiltonian for MnPSe$_3$ more straightforward. Namely, MnPS$_3$ has a distorted hexagonal 2D layer which leads to further exchange interactions as compared to MnPSe$_3$, which has identical bond lengths in the layer.

The inelastic measurements presented revealed no spin-gap (Fig.~\ref{Fig_MnPSe3_Low_E_1dcuts}) and consequently we do not consider any anisotropy in the Hamiltonian. This is consistent with the expected reduced single-ion anisotropy from the less distorted octahedral environment of the Mn ion. Moreover, the magnetic structure of spins confined to the ab-plane indicates reduced anisotropy. Our initial model was based on a 2D Hamiltonian on a honeycomb lattice,

\begin{eqnarray*}
	\mathcal{H}=\sum_{\langle ij \rangle}J_{ij} \mathbf{S}_i\cdot\mathbf{S}_j = \frac{1}{2} \sum_{ i \ne j}J_{ij} \mathbf{S}_i\cdot\mathbf{S}_j
\end{eqnarray*}

that included $J_{1ab}$, $J_{2ab}$ and $J_{3ab}$ intralayer exchange interactions shown in  Fig.~\ref{MnPSe3_INS_models}(a). The $\frac{1}{2}$ prefactor has been included in the exchange interactions to be consistent with reported parameters in the literature \cite{PhysRevB.91.235425,C9RA09030D,PhysRevB.94.184428} as well as the full spin (S=5/2). This 2D model, however, did not reproduce the observed INS scattering in MnPSe$_3$ at low energy. Therefore we extended the Hamiltonian to include interlayer exchange interactions ($J_c$) to give a 3D model isotropic Hamiltonian.

We begin with a confirmation that the previous reported results for MnPS$_3$ in Ref.~\onlinecite{WildesMnPS3} can be reproduced by the powder data.  These measurements make the approximation of a higher symmetry in which the layers are undistorted with $J_1$=0.77 meV, $J_2$=0.07 meV, $J_3$=0.18 meV and $J_c$=-0.0038 meV. An easy-axis single ion anisotropy term of 0.0086 meV is also included to model the observed spin-gap in MnPS$_3$ \cite{WildesMnPS3}. Figure \ref{MnPS3_INS_models} shows the powder data are well reproduced using these parameters.   

We now turn to modeling the exchange interactions in MnPSe$_3$, that have not been experimentally determined with INS. Exchange interactions based on a $J_{1ab}$-$J_{2ab}$-$J_{3ab}$ 2D model without an interlayer $J_c$ or anisotropy terms have been theoretically reported in several studies  \cite{PhysRevB.91.235425,C9RA09030D,PhysRevB.94.184428,ISI:000433428500001}. These considered the required antiferromagnetic  structure of MnPSe$_3$ on the 2D hexagonal lattice with the hierarchy of $J_1$$>$$J_3$$>$$J_2$ and all interactions positive (antiferromagnetic). The different studies show broad agreement. Therefore after initial confirmation of these parameters, the focus of this study will be on the interlayer interactions, $J_c$. 

Starting with the values reported in Refs.~\onlinecite{PhysRevB.91.235425, C9RA09030D} gives very close agreement to the measured excitation energy range, within 1.5$\%$. We therefore applied a scale factor and kept these values fixed in the subsequent analysis at $J_{1ab}$=0.45 meV, $J_{2ab}$=0.03 meV and $J_{3ab}$=0.19 meV. To model the data, we compared measured and calculated S($|$Q$|$, $\omega$), as well as constant energy and momentum cuts. Simulations were performed with SpinW \cite{spinW} using 100000 random orientations per Q value.  To rule out any single-ion anisotropy term being able to model the data, we applied this to a modified Hamiltonian; however, the results did not show agreement. Therefore we applied the $J_c$ term and performed least squares fitting to find the best-fit $J_c$ parameter to the data. The results are shown in  Fig.~\ref{MnPSe3_INS_models}. The measured INS data are best modeled with exchange interactions $J_{1ab}$=0.45 meV, $J_{2ab}$=0.03 meV, $J_{3ab}$=0.19 meV and $J_c$=0.031(5)meV.

Comparing the exchange interactions we have determined for MnPSe$_3$ with those measured for MnPS$_3$ reveals an enhanced  interlayer interaction energy in  MnPSe$_3$ of an order of magnitude greater than reported for MnPS$_3$ from INS \cite{WildesMnPS3}. This is coupled to a reduced energy scale for intralayer $J_{1ab}$-$J_{2ab}$-$J_{3ab}$ interactions in MnPSe$_3$ compared to MnPS$_3$. While any anisotropy present in MnPSe$_3$ is beyond the limits of our measurement, adding single ion anisotropy to the spin Hamiltonian indicates that the energy is likely on the same scale or smaller than that measured in MnPS$_3$. This suggests that interlayer coupling plays a more significant role than anisotropy in MnPSe$_3$ compared to MnPS$_3$. 

The extraction of a model spin Hamiltonian that best fits the INS data reveals MnPSe$_3$ as having dominant intralayer interactions with non-negligible interlayer interactions. Conversely, MnPS$_3$ is closer to the 2D limit in the bulk compound. This balance of competing interactions can explain the similar magnetic ordering temperatures, with lower intralayer exchange interactions in MnPSe$_3$ that drive a lower ordering temperature being offset by higher interlayer exchange interactions that stabilize long range three dimensional order in the bulk. Further theoretical and experimental investigations of MnPSe$_3$ from the bulk down to the monolayer, where $J_c=0$, will be of interest to follow the crossover from the weakly three dimensional exchange interaction model to purely 2D in-plane interactions.

\section{Conclusions}

Neutron powder diffraction and inelastic neutron scattering measurements on MnPSe$_3$ allowed model magnetic exchange interactions to be  extracted. A lack of observable symmetry lowering of the crystal structure or spin-gap in the inelastic excitation spectra is consistent with magnetic order confined to the $ab$-plane. The measured magnetic excitations show well-defined spin waves that were modeled within a linear spin wave theory. While the intralayer  $J_{1ab}$-$J_{2ab}$-$J_{3ab}$ interactions dominate the magnetic behavior, extension beyond a 2D Hamiltonian was required with the inclusion of a non-negligible interlayer interaction $J_c=0.031(5)$ meV to model the data.

\begin{acknowledgments}
This research used resources at the High Flux Isotope Reactor and Spallation Neutron Source, a DOE Office of Science User Facility operated by the Oak Ridge National Laboratory. D.M. acknowledges support from the Gordon and Betty Moore Foundation’s EPiQS Initiative, Grant GBMF9069. This manuscript has been authored by UT-Battelle, LLC under Contract No. DE-AC05-00OR22725 with the U.S. Department of Energy. The United States Government retains and the publisher, by accepting the article for publication, acknowledges that the United States Government retains a non-exclusive, paidup, irrevocable, world-wide license to publish or reproduce the published form of this manuscript, or allow others to do so, for United States Government purposes. The Department of Energy will provide public access to these results of federally sponsored research in accordance with the DOE Public Access Plan(http://energy.gov/downloads/doepublic-access-plan).
\end{acknowledgments}


\begin{thebibliography}{63}%
	\makeatletter
	\providecommand \@ifxundefined [1]{%
		\@ifx{#1\undefined}
	}%
	\providecommand \@ifnum [1]{%
		\ifnum #1\expandafter \@firstoftwo
		\else \expandafter \@secondoftwo
		\fi
	}%
	\providecommand \@ifx [1]{%
		\ifx #1\expandafter \@firstoftwo
		\else \expandafter \@secondoftwo
		\fi
	}%
	\providecommand \natexlab [1]{#1}%
	\providecommand \enquote  [1]{``#1''}%
	\providecommand \bibnamefont  [1]{#1}%
	\providecommand \bibfnamefont [1]{#1}%
	\providecommand \citenamefont [1]{#1}%
	\providecommand \href@noop [0]{\@secondoftwo}%
	\providecommand \href [0]{\begingroup \@sanitize@url \@href}%
	\providecommand \@href[1]{\@@startlink{#1}\@@href}%
	\providecommand \@@href[1]{\endgroup#1\@@endlink}%
	\providecommand \@sanitize@url [0]{\catcode `\\12\catcode `\$12\catcode
		`\&12\catcode `\#12\catcode `\^12\catcode `\_12\catcode `\%12\relax}%
	\providecommand \@@startlink[1]{}%
	\providecommand \@@endlink[0]{}%
	\providecommand \url  [0]{\begingroup\@sanitize@url \@url }%
	\providecommand \@url [1]{\endgroup\@href {#1}{\urlprefix }}%
	\providecommand \urlprefix  [0]{URL }%
	\providecommand \Eprint [0]{\href }%
	\providecommand \doibase [0]{http://dx.doi.org/}%
	\providecommand \selectlanguage [0]{\@gobble}%
	\providecommand \bibinfo  [0]{\@secondoftwo}%
	\providecommand \bibfield  [0]{\@secondoftwo}%
	\providecommand \translation [1]{[#1]}%
	\providecommand \BibitemOpen [0]{}%
	\providecommand \bibitemStop [0]{}%
	\providecommand \bibitemNoStop [0]{.\EOS\space}%
	\providecommand \EOS [0]{\spacefactor3000\relax}%
	\providecommand \BibitemShut  [1]{\csname bibitem#1\endcsname}%
	\let\auto@bib@innerbib\@empty
	\bibitem [{\citenamefont {Novoselov}\ \emph {et~al.}(2005)\citenamefont
		{Novoselov}, \citenamefont {Geim}, \citenamefont {Morozov}, \citenamefont
		{Jiang}, \citenamefont {Katsnelson}, \citenamefont {Grigorieva},
		\citenamefont {Dubonos},\ and\ \citenamefont {Firsov}}]{NatureGraphene}%
	\BibitemOpen
	\bibfield  {author} {\bibinfo {author} {\bibfnamefont {K.~S.}\ \bibnamefont
			{Novoselov}}, \bibinfo {author} {\bibfnamefont {A.~K.}\ \bibnamefont {Geim}},
		\bibinfo {author} {\bibfnamefont {S.~V.}\ \bibnamefont {Morozov}}, \bibinfo
		{author} {\bibfnamefont {D.}~\bibnamefont {Jiang}}, \bibinfo {author}
		{\bibfnamefont {M.~I.}\ \bibnamefont {Katsnelson}}, \bibinfo {author}
		{\bibfnamefont {I.~V.}\ \bibnamefont {Grigorieva}}, \bibinfo {author}
		{\bibfnamefont {S.~V.}\ \bibnamefont {Dubonos}}, \ and\ \bibinfo {author}
		{\bibfnamefont {A.~A.}\ \bibnamefont {Firsov}},\ }\href@noop {} {\bibfield
		{journal} {\bibinfo  {journal} {Nature}\ }\textbf {\bibinfo {volume} {438}},\
		\bibinfo {pages} {197} (\bibinfo {year} {2005})}\BibitemShut {NoStop}%
	\bibitem [{\citenamefont {Zhang}\ \emph {et~al.}(2005)\citenamefont {Zhang},
		\citenamefont {Tan}, \citenamefont {Stormer},\ and\ \citenamefont
		{Kim}}]{ZhangQHallGraphene}%
	\BibitemOpen
	\bibfield  {author} {\bibinfo {author} {\bibfnamefont {Y.~B.}\ \bibnamefont
			{Zhang}}, \bibinfo {author} {\bibfnamefont {Y.~W.}\ \bibnamefont {Tan}},
		\bibinfo {author} {\bibfnamefont {H.~L.}\ \bibnamefont {Stormer}}, \ and\
		\bibinfo {author} {\bibfnamefont {P.}~\bibnamefont {Kim}},\ }\href@noop {}
	{\bibfield  {journal} {\bibinfo  {journal} {Nature}\ }\textbf {\bibinfo
			{volume} {438}},\ \bibinfo {pages} {201} (\bibinfo {year}
		{2005})}\BibitemShut {NoStop}%
	\bibitem [{\citenamefont {Yu}\ \emph {et~al.}(2010)\citenamefont {Yu},
		\citenamefont {Zhang}, \citenamefont {Zhang}, \citenamefont {Zhang},
		\citenamefont {Dai},\ and\ \citenamefont {Fang}}]{Yu61}%
	\BibitemOpen
	\bibfield  {author} {\bibinfo {author} {\bibfnamefont {R.}~\bibnamefont
			{Yu}}, \bibinfo {author} {\bibfnamefont {W.}~\bibnamefont {Zhang}}, \bibinfo
		{author} {\bibfnamefont {H.-J.}\ \bibnamefont {Zhang}}, \bibinfo {author}
		{\bibfnamefont {S.-C.}\ \bibnamefont {Zhang}}, \bibinfo {author}
		{\bibfnamefont {X.}~\bibnamefont {Dai}}, \ and\ \bibinfo {author}
		{\bibfnamefont {Z.}~\bibnamefont {Fang}},\ }\href@noop {} {\bibfield
		{journal} {\bibinfo  {journal} {Science}\ }\textbf {\bibinfo {volume}
			{329}},\ \bibinfo {pages} {61} (\bibinfo {year} {2010})}\BibitemShut
	{NoStop}%
	\bibitem [{\citenamefont {Bhimanapati}\ \emph {et~al.}(2015)\citenamefont
		{Bhimanapati}, \citenamefont {Lin}, \citenamefont {Meunier}, \citenamefont
		{Jung}, \citenamefont {Cha}, \citenamefont {Das}, \citenamefont {Xiao},
		\citenamefont {Son}, \citenamefont {Strano}, \citenamefont {Cooper},
		\citenamefont {Liang}, \citenamefont {Louie}, \citenamefont {Ringe},
		\citenamefont {Zhou}, \citenamefont {Kim}, \citenamefont {Naik},
		\citenamefont {Sumpter}, \citenamefont {Terrones}, \citenamefont {Xia},
		\citenamefont {Wang}, \citenamefont {Zhu}, \citenamefont {Akinwande},
		\citenamefont {Alem}, \citenamefont {Schuller}, \citenamefont {Schaak},
		\citenamefont {Terrones},\ and\ \citenamefont
		{Robinson}}]{doi:10.1021/acsnano.5b05556}%
	\BibitemOpen
	\bibfield  {author} {\bibinfo {author} {\bibfnamefont {G.~R.}\ \bibnamefont
			{Bhimanapati}}, \bibinfo {author} {\bibfnamefont {Z.}~\bibnamefont {Lin}},
		\bibinfo {author} {\bibfnamefont {V.}~\bibnamefont {Meunier}}, \bibinfo
		{author} {\bibfnamefont {Y.}~\bibnamefont {Jung}}, \bibinfo {author}
		{\bibfnamefont {J.}~\bibnamefont {Cha}}, \bibinfo {author} {\bibfnamefont
			{S.}~\bibnamefont {Das}}, \bibinfo {author} {\bibfnamefont {D.}~\bibnamefont
			{Xiao}}, \bibinfo {author} {\bibfnamefont {Y.}~\bibnamefont {Son}}, \bibinfo
		{author} {\bibfnamefont {M.~S.}\ \bibnamefont {Strano}}, \bibinfo {author}
		{\bibfnamefont {V.~R.}\ \bibnamefont {Cooper}}, \bibinfo {author}
		{\bibfnamefont {L.}~\bibnamefont {Liang}}, \bibinfo {author} {\bibfnamefont
			{S.~G.}\ \bibnamefont {Louie}}, \bibinfo {author} {\bibfnamefont
			{E.}~\bibnamefont {Ringe}}, \bibinfo {author} {\bibfnamefont
			{W.}~\bibnamefont {Zhou}}, \bibinfo {author} {\bibfnamefont {S.~S.}\
			\bibnamefont {Kim}}, \bibinfo {author} {\bibfnamefont {R.~R.}\ \bibnamefont
			{Naik}}, \bibinfo {author} {\bibfnamefont {B.~G.}\ \bibnamefont {Sumpter}},
		\bibinfo {author} {\bibfnamefont {H.}~\bibnamefont {Terrones}}, \bibinfo
		{author} {\bibfnamefont {F.}~\bibnamefont {Xia}}, \bibinfo {author}
		{\bibfnamefont {Y.}~\bibnamefont {Wang}}, \bibinfo {author} {\bibfnamefont
			{J.}~\bibnamefont {Zhu}}, \bibinfo {author} {\bibfnamefont {D.}~\bibnamefont
			{Akinwande}}, \bibinfo {author} {\bibfnamefont {N.}~\bibnamefont {Alem}},
		\bibinfo {author} {\bibfnamefont {J.~A.}\ \bibnamefont {Schuller}}, \bibinfo
		{author} {\bibfnamefont {R.~E.}\ \bibnamefont {Schaak}}, \bibinfo {author}
		{\bibfnamefont {M.}~\bibnamefont {Terrones}}, \ and\ \bibinfo {author}
		{\bibfnamefont {J.~A.}\ \bibnamefont {Robinson}},\ }\href {\doibase
		10.1021/acsnano.5b05556} {\bibfield  {journal} {\bibinfo  {journal} {ACS
				Nano}\ }\textbf {\bibinfo {volume} {9}},\ \bibinfo {pages} {11509} (\bibinfo
		{year} {2015})}\BibitemShut {NoStop}%
	\bibitem [{\citenamefont {Mermin}\ and\ \citenamefont
		{Wagner}(1966)}]{PhysRevLett.17.1133}%
	\BibitemOpen
	\bibfield  {author} {\bibinfo {author} {\bibfnamefont {N.~D.}\ \bibnamefont
			{Mermin}}\ and\ \bibinfo {author} {\bibfnamefont {H.}~\bibnamefont
			{Wagner}},\ }\href {\doibase 10.1103/PhysRevLett.17.1133} {\bibfield
		{journal} {\bibinfo  {journal} {Phys. Rev. Lett.}\ }\textbf {\bibinfo
			{volume} {17}},\ \bibinfo {pages} {1133} (\bibinfo {year}
		{1966})}\BibitemShut {NoStop}%
	\bibitem [{\citenamefont {Onsager}(1944)}]{PhysRev.65.117}%
	\BibitemOpen
	\bibfield  {author} {\bibinfo {author} {\bibfnamefont {L.}~\bibnamefont
			{Onsager}},\ }\href {\doibase 10.1103/PhysRev.65.117} {\bibfield  {journal}
		{\bibinfo  {journal} {Phys. Rev.}\ }\textbf {\bibinfo {volume} {65}},\
		\bibinfo {pages} {117} (\bibinfo {year} {1944})}\BibitemShut {NoStop}%
	\bibitem [{\citenamefont {Deiseroth}\ \emph {et~al.}(2006)\citenamefont
		{Deiseroth}, \citenamefont {Aleksandrov}, \citenamefont {Reiner},
		\citenamefont {Kienle},\ and\ \citenamefont
		{Kremer}}]{FGTdoi:10.1002/ejic.200501020}%
	\BibitemOpen
	\bibfield  {author} {\bibinfo {author} {\bibfnamefont {H.-J.}\ \bibnamefont
			{Deiseroth}}, \bibinfo {author} {\bibfnamefont {K.}~\bibnamefont
			{Aleksandrov}}, \bibinfo {author} {\bibfnamefont {C.}~\bibnamefont {Reiner}},
		\bibinfo {author} {\bibfnamefont {L.}~\bibnamefont {Kienle}}, \ and\ \bibinfo
		{author} {\bibfnamefont {R.~K.}\ \bibnamefont {Kremer}},\ }\href@noop {}
	{\bibfield  {journal} {\bibinfo  {journal} {European Journal of Inorganic
				Chemistry}\ }\textbf {\bibinfo {volume} {2006}},\ \bibinfo {pages} {1561}
		(\bibinfo {year} {2006})}\BibitemShut {NoStop}%
	\bibitem [{\citenamefont {May}\ \emph {et~al.}(2016)\citenamefont {May},
		\citenamefont {Calder}, \citenamefont {Cantoni}, \citenamefont {Cao},\ and\
		\citenamefont {McGuire}}]{PhysRevB.93.014411}%
	\BibitemOpen
	\bibfield  {author} {\bibinfo {author} {\bibfnamefont {A.~F.}\ \bibnamefont
			{May}}, \bibinfo {author} {\bibfnamefont {S.}~\bibnamefont {Calder}},
		\bibinfo {author} {\bibfnamefont {C.}~\bibnamefont {Cantoni}}, \bibinfo
		{author} {\bibfnamefont {H.}~\bibnamefont {Cao}}, \ and\ \bibinfo {author}
		{\bibfnamefont {M.~A.}\ \bibnamefont {McGuire}},\ }\href {\doibase
		10.1103/PhysRevB.93.014411} {\bibfield  {journal} {\bibinfo  {journal} {Phys.
				Rev. B}\ }\textbf {\bibinfo {volume} {93}},\ \bibinfo {pages} {014411}
		(\bibinfo {year} {2016})}\BibitemShut {NoStop}%
	\bibitem [{\citenamefont {Fei}\ \emph {et~al.}(2018)\citenamefont {Fei},
		\citenamefont {Huang}, \citenamefont {Malinowski}, \citenamefont {Wang},
		\citenamefont {Song}, \citenamefont {Sanchez}, \citenamefont {Yao},
		\citenamefont {Xiao}, \citenamefont {Zhu}, \citenamefont {May}, \citenamefont
		{Wu}, \citenamefont {Cobden}, \citenamefont {Chu},\ and\ \citenamefont
		{Xu}}]{ISI:000442526400011}%
	\BibitemOpen
	\bibfield  {author} {\bibinfo {author} {\bibfnamefont {Z.}~\bibnamefont
			{Fei}}, \bibinfo {author} {\bibfnamefont {B.}~\bibnamefont {Huang}}, \bibinfo
		{author} {\bibfnamefont {P.}~\bibnamefont {Malinowski}}, \bibinfo {author}
		{\bibfnamefont {W.}~\bibnamefont {Wang}}, \bibinfo {author} {\bibfnamefont
			{T.}~\bibnamefont {Song}}, \bibinfo {author} {\bibfnamefont {J.}~\bibnamefont
			{Sanchez}}, \bibinfo {author} {\bibfnamefont {W.}~\bibnamefont {Yao}},
		\bibinfo {author} {\bibfnamefont {D.}~\bibnamefont {Xiao}}, \bibinfo {author}
		{\bibfnamefont {X.}~\bibnamefont {Zhu}}, \bibinfo {author} {\bibfnamefont
			{A.~F.}\ \bibnamefont {May}}, \bibinfo {author} {\bibfnamefont
			{W.}~\bibnamefont {Wu}}, \bibinfo {author} {\bibfnamefont {D.~H.}\
			\bibnamefont {Cobden}}, \bibinfo {author} {\bibfnamefont {J.-H.}\
			\bibnamefont {Chu}}, \ and\ \bibinfo {author} {\bibfnamefont
			{X.}~\bibnamefont {Xu}},\ }\href@noop {} {\bibfield  {journal} {\bibinfo
			{journal} {Nature Materials}\ }\textbf {\bibinfo {volume} {17}},\ \bibinfo
		{pages} {778} (\bibinfo {year} {2018})}\BibitemShut {NoStop}%
	\bibitem [{\citenamefont {Tan}\ \emph {et~al.}(2018)\citenamefont {Tan},
		\citenamefont {Lee}, \citenamefont {Jung}, \citenamefont {Park},
		\citenamefont {Albarakati}, \citenamefont {Partridge}, \citenamefont {Field},
		\citenamefont {McCulloch}, \citenamefont {Wang},\ and\ \citenamefont
		{Lee}}]{ISI:000430389100006}%
	\BibitemOpen
	\bibfield  {author} {\bibinfo {author} {\bibfnamefont {C.}~\bibnamefont
			{Tan}}, \bibinfo {author} {\bibfnamefont {J.}~\bibnamefont {Lee}}, \bibinfo
		{author} {\bibfnamefont {S.-G.}\ \bibnamefont {Jung}}, \bibinfo {author}
		{\bibfnamefont {T.}~\bibnamefont {Park}}, \bibinfo {author} {\bibfnamefont
			{S.}~\bibnamefont {Albarakati}}, \bibinfo {author} {\bibfnamefont
			{J.}~\bibnamefont {Partridge}}, \bibinfo {author} {\bibfnamefont {M.~R.}\
			\bibnamefont {Field}}, \bibinfo {author} {\bibfnamefont {D.~G.}\ \bibnamefont
			{McCulloch}}, \bibinfo {author} {\bibfnamefont {L.}~\bibnamefont {Wang}}, \
		and\ \bibinfo {author} {\bibfnamefont {C.}~\bibnamefont {Lee}},\ }\href@noop
	{} {\bibfield  {journal} {\bibinfo  {journal} {Nature Communications}\
		}\textbf {\bibinfo {volume} {{9}}} (\bibinfo {year} {{2018}})}\BibitemShut
	{NoStop}%
	\bibitem [{\citenamefont {Johansen}\ \emph {et~al.}(2019)\citenamefont
		{Johansen}, \citenamefont {Risingg\aa{}rd}, \citenamefont {Sudb\o{}},
		\citenamefont {Linder},\ and\ \citenamefont
		{Brataas}}]{PhysRevLett.122.217203}%
	\BibitemOpen
	\bibfield  {author} {\bibinfo {author} {\bibfnamefont {O.}~\bibnamefont
			{Johansen}}, \bibinfo {author} {\bibfnamefont {V.}~\bibnamefont
			{Risingg\aa{}rd}}, \bibinfo {author} {\bibfnamefont {A.}~\bibnamefont
			{Sudb\o{}}}, \bibinfo {author} {\bibfnamefont {J.}~\bibnamefont {Linder}}, \
		and\ \bibinfo {author} {\bibfnamefont {A.}~\bibnamefont {Brataas}},\
	}\href@noop {} {\bibfield  {journal} {\bibinfo  {journal} {Phys. Rev. Lett.}\
		}\textbf {\bibinfo {volume} {122}},\ \bibinfo {pages} {217203} (\bibinfo
		{year} {2019})}\BibitemShut {NoStop}%
	\bibitem [{\citenamefont {Calder}\ \emph {et~al.}(2019)\citenamefont {Calder},
		\citenamefont {Kolesnikov},\ and\ \citenamefont {May}}]{PhysRevB.99.094423}%
	\BibitemOpen
	\bibfield  {author} {\bibinfo {author} {\bibfnamefont {S.}~\bibnamefont
			{Calder}}, \bibinfo {author} {\bibfnamefont {A.~I.}\ \bibnamefont
			{Kolesnikov}}, \ and\ \bibinfo {author} {\bibfnamefont {A.~F.}\ \bibnamefont
			{May}},\ }\href@noop {} {\bibfield  {journal} {\bibinfo  {journal} {Phys.
				Rev. B}\ }\textbf {\bibinfo {volume} {99}},\ \bibinfo {pages} {094423}
		(\bibinfo {year} {2019})}\BibitemShut {NoStop}%
	\bibitem [{\citenamefont {Wang}\ \emph {et~al.}(2019)\citenamefont {Wang},
		\citenamefont {Tang}, \citenamefont {Xia}, \citenamefont {He}, \citenamefont
		{Zhang}, \citenamefont {Liu}, \citenamefont {Wan}, \citenamefont {Fang},
		\citenamefont {Guo}, \citenamefont {Yang}, \citenamefont {Guang},
		\citenamefont {Zhang}, \citenamefont {Xu}, \citenamefont {Wei}, \citenamefont
		{Liao}, \citenamefont {Lu}, \citenamefont {Feng}, \citenamefont {Li},
		\citenamefont {Peng}, \citenamefont {Wei}, \citenamefont {Yang},
		\citenamefont {Shi}, \citenamefont {Zhang}, \citenamefont {Han},
		\citenamefont {Zhang}, \citenamefont {Zhang}, \citenamefont {Yu},\ and\
		\citenamefont {Han}}]{Wangeaaw8904}%
	\BibitemOpen
	\bibfield  {author} {\bibinfo {author} {\bibfnamefont {X.}~\bibnamefont
			{Wang}}, \bibinfo {author} {\bibfnamefont {J.}~\bibnamefont {Tang}}, \bibinfo
		{author} {\bibfnamefont {X.}~\bibnamefont {Xia}}, \bibinfo {author}
		{\bibfnamefont {C.}~\bibnamefont {He}}, \bibinfo {author} {\bibfnamefont
			{J.}~\bibnamefont {Zhang}}, \bibinfo {author} {\bibfnamefont
			{Y.}~\bibnamefont {Liu}}, \bibinfo {author} {\bibfnamefont {C.}~\bibnamefont
			{Wan}}, \bibinfo {author} {\bibfnamefont {C.}~\bibnamefont {Fang}}, \bibinfo
		{author} {\bibfnamefont {C.}~\bibnamefont {Guo}}, \bibinfo {author}
		{\bibfnamefont {W.}~\bibnamefont {Yang}}, \bibinfo {author} {\bibfnamefont
			{Y.}~\bibnamefont {Guang}}, \bibinfo {author} {\bibfnamefont
			{X.}~\bibnamefont {Zhang}}, \bibinfo {author} {\bibfnamefont
			{H.}~\bibnamefont {Xu}}, \bibinfo {author} {\bibfnamefont {J.}~\bibnamefont
			{Wei}}, \bibinfo {author} {\bibfnamefont {M.}~\bibnamefont {Liao}}, \bibinfo
		{author} {\bibfnamefont {X.}~\bibnamefont {Lu}}, \bibinfo {author}
		{\bibfnamefont {J.}~\bibnamefont {Feng}}, \bibinfo {author} {\bibfnamefont
			{X.}~\bibnamefont {Li}}, \bibinfo {author} {\bibfnamefont {Y.}~\bibnamefont
			{Peng}}, \bibinfo {author} {\bibfnamefont {H.}~\bibnamefont {Wei}}, \bibinfo
		{author} {\bibfnamefont {R.}~\bibnamefont {Yang}}, \bibinfo {author}
		{\bibfnamefont {D.}~\bibnamefont {Shi}}, \bibinfo {author} {\bibfnamefont
			{X.}~\bibnamefont {Zhang}}, \bibinfo {author} {\bibfnamefont
			{Z.}~\bibnamefont {Han}}, \bibinfo {author} {\bibfnamefont {Z.}~\bibnamefont
			{Zhang}}, \bibinfo {author} {\bibfnamefont {G.}~\bibnamefont {Zhang}},
		\bibinfo {author} {\bibfnamefont {G.}~\bibnamefont {Yu}}, \ and\ \bibinfo
		{author} {\bibfnamefont {X.}~\bibnamefont {Han}},\ }\href@noop {} {\bibfield
		{journal} {\bibinfo  {journal} {Science Advances}\ }\textbf {\bibinfo
			{volume} {5}} (\bibinfo {year} {2019})}\BibitemShut {NoStop}%
	\bibitem [{\citenamefont {Carteaux}\ \emph {et~al.}(1995)\citenamefont
		{Carteaux}, \citenamefont {Moussa},\ and\ \citenamefont
		{Spiesser}}]{Carteaux_1995}%
	\BibitemOpen
	\bibfield  {author} {\bibinfo {author} {\bibfnamefont {V.}~\bibnamefont
			{Carteaux}}, \bibinfo {author} {\bibfnamefont {F.}~\bibnamefont {Moussa}}, \
		and\ \bibinfo {author} {\bibfnamefont {M.}~\bibnamefont {Spiesser}},\
	}\href@noop {} {\bibfield  {journal} {\bibinfo  {journal} {Europhysics
				Letters ({EPL})}\ }\textbf {\bibinfo {volume} {29}},\ \bibinfo {pages} {251}
		(\bibinfo {year} {1995})}\BibitemShut {NoStop}%
	\bibitem [{\citenamefont {Williams}\ \emph {et~al.}(2015)\citenamefont
		{Williams}, \citenamefont {Aczel}, \citenamefont {Lumsden}, \citenamefont
		{Nagler}, \citenamefont {Stone}, \citenamefont {Yan},\ and\ \citenamefont
		{Mandrus}}]{PhysRevB.92.144404}%
	\BibitemOpen
	\bibfield  {author} {\bibinfo {author} {\bibfnamefont {T.~J.}\ \bibnamefont
			{Williams}}, \bibinfo {author} {\bibfnamefont {A.~A.}\ \bibnamefont {Aczel}},
		\bibinfo {author} {\bibfnamefont {M.~D.}\ \bibnamefont {Lumsden}}, \bibinfo
		{author} {\bibfnamefont {S.~E.}\ \bibnamefont {Nagler}}, \bibinfo {author}
		{\bibfnamefont {M.~B.}\ \bibnamefont {Stone}}, \bibinfo {author}
		{\bibfnamefont {J.-Q.}\ \bibnamefont {Yan}}, \ and\ \bibinfo {author}
		{\bibfnamefont {D.}~\bibnamefont {Mandrus}},\ }\href {\doibase
		10.1103/PhysRevB.92.144404} {\bibfield  {journal} {\bibinfo  {journal} {Phys.
				Rev. B}\ }\textbf {\bibinfo {volume} {92}},\ \bibinfo {pages} {144404}
		(\bibinfo {year} {2015})}\BibitemShut {NoStop}%
	\bibitem [{\citenamefont {Zhang}\ \emph {et~al.}(2019)\citenamefont {Zhang},
		\citenamefont {Cai}, \citenamefont {Xia}, \citenamefont {Liang},
		\citenamefont {Huang}, \citenamefont {Wang}, \citenamefont {Yang},
		\citenamefont {Yuan}, \citenamefont {Chen}, \citenamefont {Zhang},
		\citenamefont {Guo}, \citenamefont {Liu},\ and\ \citenamefont
		{Li}}]{PhysRevLett.123.047203}%
	\BibitemOpen
	\bibfield  {author} {\bibinfo {author} {\bibfnamefont {J.}~\bibnamefont
			{Zhang}}, \bibinfo {author} {\bibfnamefont {X.}~\bibnamefont {Cai}}, \bibinfo
		{author} {\bibfnamefont {W.}~\bibnamefont {Xia}}, \bibinfo {author}
		{\bibfnamefont {A.}~\bibnamefont {Liang}}, \bibinfo {author} {\bibfnamefont
			{J.}~\bibnamefont {Huang}}, \bibinfo {author} {\bibfnamefont
			{C.}~\bibnamefont {Wang}}, \bibinfo {author} {\bibfnamefont {L.}~\bibnamefont
			{Yang}}, \bibinfo {author} {\bibfnamefont {H.}~\bibnamefont {Yuan}}, \bibinfo
		{author} {\bibfnamefont {Y.}~\bibnamefont {Chen}}, \bibinfo {author}
		{\bibfnamefont {S.}~\bibnamefont {Zhang}}, \bibinfo {author} {\bibfnamefont
			{Y.}~\bibnamefont {Guo}}, \bibinfo {author} {\bibfnamefont {Z.}~\bibnamefont
			{Liu}}, \ and\ \bibinfo {author} {\bibfnamefont {G.}~\bibnamefont {Li}},\
	}\href@noop {} {\bibfield  {journal} {\bibinfo  {journal} {Phys. Rev. Lett.}\
		}\textbf {\bibinfo {volume} {123}},\ \bibinfo {pages} {047203} (\bibinfo
		{year} {2019})}\BibitemShut {NoStop}%
	\bibitem [{\citenamefont {Ito}\ \emph {et~al.}(2019)\citenamefont {Ito},
		\citenamefont {Kikkawa}, \citenamefont {Barker}, \citenamefont {Hirobe},
		\citenamefont {Shiomi},\ and\ \citenamefont {Saitoh}}]{PhysRevB.100.060402}%
	\BibitemOpen
	\bibfield  {author} {\bibinfo {author} {\bibfnamefont {N.}~\bibnamefont
			{Ito}}, \bibinfo {author} {\bibfnamefont {T.}~\bibnamefont {Kikkawa}},
		\bibinfo {author} {\bibfnamefont {J.}~\bibnamefont {Barker}}, \bibinfo
		{author} {\bibfnamefont {D.}~\bibnamefont {Hirobe}}, \bibinfo {author}
		{\bibfnamefont {Y.}~\bibnamefont {Shiomi}}, \ and\ \bibinfo {author}
		{\bibfnamefont {E.}~\bibnamefont {Saitoh}},\ }\href@noop {} {\bibfield
		{journal} {\bibinfo  {journal} {Phys. Rev. B}\ }\textbf {\bibinfo {volume}
			{100}},\ \bibinfo {pages} {060402} (\bibinfo {year} {2019})}\BibitemShut
	{NoStop}%
	\bibitem [{\citenamefont {Louisy}\ \emph {et~al.}(1978)\citenamefont {Louisy},
		\citenamefont {Ouvrard}, \citenamefont {Schleich},\ and\ \citenamefont
		{Brec}}]{LOUISY197861}%
	\BibitemOpen
	\bibfield  {author} {\bibinfo {author} {\bibfnamefont {A.}~\bibnamefont
			{Louisy}}, \bibinfo {author} {\bibfnamefont {G.}~\bibnamefont {Ouvrard}},
		\bibinfo {author} {\bibfnamefont {D.}~\bibnamefont {Schleich}}, \ and\
		\bibinfo {author} {\bibfnamefont {R.}~\bibnamefont {Brec}},\ }\href {\doibase
		https://doi.org/10.1016/0038-1098(78)90328-9} {\bibfield  {journal} {\bibinfo
			{journal} {Solid State Communications}\ }\textbf {\bibinfo {volume} {28}},\
		\bibinfo {pages} {61 } (\bibinfo {year} {1978})}\BibitemShut {NoStop}%
	\bibitem [{\citenamefont {Calder}\ \emph {et~al.}(2020)\citenamefont {Calder},
		\citenamefont {Haglund}, \citenamefont {Liu}, \citenamefont {Pajerowski},
		\citenamefont {Cao}, \citenamefont {Williams}, \citenamefont {Garlea},\ and\
		\citenamefont {Mandrus}}]{PhysRevB.102.024408}%
	\BibitemOpen
	\bibfield  {author} {\bibinfo {author} {\bibfnamefont {S.}~\bibnamefont
			{Calder}}, \bibinfo {author} {\bibfnamefont {A.~V.}\ \bibnamefont {Haglund}},
		\bibinfo {author} {\bibfnamefont {Y.}~\bibnamefont {Liu}}, \bibinfo {author}
		{\bibfnamefont {D.~M.}\ \bibnamefont {Pajerowski}}, \bibinfo {author}
		{\bibfnamefont {H.~B.}\ \bibnamefont {Cao}}, \bibinfo {author} {\bibfnamefont
			{T.~J.}\ \bibnamefont {Williams}}, \bibinfo {author} {\bibfnamefont {V.~O.}\
			\bibnamefont {Garlea}}, \ and\ \bibinfo {author} {\bibfnamefont
			{D.}~\bibnamefont {Mandrus}},\ }\href {\doibase 10.1103/PhysRevB.102.024408}
	{\bibfield  {journal} {\bibinfo  {journal} {Phys. Rev. B}\ }\textbf {\bibinfo
			{volume} {102}},\ \bibinfo {pages} {024408} (\bibinfo {year}
		{2020})}\BibitemShut {NoStop}%
	\bibitem [{\citenamefont {McGuire}\ \emph {et~al.}(2017)\citenamefont
		{McGuire}, \citenamefont {Clark}, \citenamefont {KC}, \citenamefont {Chance},
		\citenamefont {Jellison}, \citenamefont {Cooper}, \citenamefont {Xu},\ and\
		\citenamefont {Sales}}]{PhysRevMaterials.1.014001}%
	\BibitemOpen
	\bibfield  {author} {\bibinfo {author} {\bibfnamefont {M.~A.}\ \bibnamefont
			{McGuire}}, \bibinfo {author} {\bibfnamefont {G.}~\bibnamefont {Clark}},
		\bibinfo {author} {\bibfnamefont {S.}~\bibnamefont {KC}}, \bibinfo {author}
		{\bibfnamefont {W.~M.}\ \bibnamefont {Chance}}, \bibinfo {author}
		{\bibfnamefont {G.~E.}\ \bibnamefont {Jellison}}, \bibinfo {author}
		{\bibfnamefont {V.~R.}\ \bibnamefont {Cooper}}, \bibinfo {author}
		{\bibfnamefont {X.}~\bibnamefont {Xu}}, \ and\ \bibinfo {author}
		{\bibfnamefont {B.~C.}\ \bibnamefont {Sales}},\ }\href {\doibase
		10.1103/PhysRevMaterials.1.014001} {\bibfield  {journal} {\bibinfo  {journal}
			{Phys. Rev. Materials}\ }\textbf {\bibinfo {volume} {1}},\ \bibinfo {pages}
		{014001} (\bibinfo {year} {2017})}\BibitemShut {NoStop}%
	\bibitem [{\citenamefont {Yan}\ \emph {et~al.}(2019)\citenamefont {Yan},
		\citenamefont {Luo}, \citenamefont {Chen}, \citenamefont {Gao}, \citenamefont
		{Jiang}, \citenamefont {Zhao}, \citenamefont {Sun}, \citenamefont {Lv},
		\citenamefont {Tian}, \citenamefont {Yin}, \citenamefont {Lei}, \citenamefont
		{Lu}, \citenamefont {Tong}, \citenamefont {Song}, \citenamefont {Zhu},\ and\
		\citenamefont {Sun}}]{PhysRevB.100.094402}%
	\BibitemOpen
	\bibfield  {author} {\bibinfo {author} {\bibfnamefont {J.}~\bibnamefont
			{Yan}}, \bibinfo {author} {\bibfnamefont {X.}~\bibnamefont {Luo}}, \bibinfo
		{author} {\bibfnamefont {F.~C.}\ \bibnamefont {Chen}}, \bibinfo {author}
		{\bibfnamefont {J.~J.}\ \bibnamefont {Gao}}, \bibinfo {author} {\bibfnamefont
			{Z.~Z.}\ \bibnamefont {Jiang}}, \bibinfo {author} {\bibfnamefont {G.~C.}\
			\bibnamefont {Zhao}}, \bibinfo {author} {\bibfnamefont {Y.}~\bibnamefont
			{Sun}}, \bibinfo {author} {\bibfnamefont {H.~Y.}\ \bibnamefont {Lv}},
		\bibinfo {author} {\bibfnamefont {S.~J.}\ \bibnamefont {Tian}}, \bibinfo
		{author} {\bibfnamefont {Q.~W.}\ \bibnamefont {Yin}}, \bibinfo {author}
		{\bibfnamefont {H.~C.}\ \bibnamefont {Lei}}, \bibinfo {author} {\bibfnamefont
			{W.~J.}\ \bibnamefont {Lu}}, \bibinfo {author} {\bibfnamefont
			{P.}~\bibnamefont {Tong}}, \bibinfo {author} {\bibfnamefont {W.~H.}\
			\bibnamefont {Song}}, \bibinfo {author} {\bibfnamefont {X.~B.}\ \bibnamefont
			{Zhu}}, \ and\ \bibinfo {author} {\bibfnamefont {Y.~P.}\ \bibnamefont
			{Sun}},\ }\href {\doibase 10.1103/PhysRevB.100.094402} {\bibfield  {journal}
		{\bibinfo  {journal} {Phys. Rev. B}\ }\textbf {\bibinfo {volume} {100}},\
		\bibinfo {pages} {094402} (\bibinfo {year} {2019})}\BibitemShut {NoStop}%
	\bibitem [{\citenamefont {Son}\ \emph {et~al.}(2019)\citenamefont {Son},
		\citenamefont {Coak}, \citenamefont {Lee}, \citenamefont {Kim}, \citenamefont
		{Kim}, \citenamefont {Hamidov}, \citenamefont {Cho}, \citenamefont {Liu},
		\citenamefont {Jarvis}, \citenamefont {Brown}, \citenamefont {Kim},
		\citenamefont {Park}, \citenamefont {Khomskii}, \citenamefont {Saxena},\ and\
		\citenamefont {Park}}]{PhysRevB.99.041402}%
	\BibitemOpen
	\bibfield  {author} {\bibinfo {author} {\bibfnamefont {S.}~\bibnamefont
			{Son}}, \bibinfo {author} {\bibfnamefont {M.~J.}\ \bibnamefont {Coak}},
		\bibinfo {author} {\bibfnamefont {N.}~\bibnamefont {Lee}}, \bibinfo {author}
		{\bibfnamefont {J.}~\bibnamefont {Kim}}, \bibinfo {author} {\bibfnamefont
			{T.~Y.}\ \bibnamefont {Kim}}, \bibinfo {author} {\bibfnamefont
			{H.}~\bibnamefont {Hamidov}}, \bibinfo {author} {\bibfnamefont
			{H.}~\bibnamefont {Cho}}, \bibinfo {author} {\bibfnamefont {C.}~\bibnamefont
			{Liu}}, \bibinfo {author} {\bibfnamefont {D.~M.}\ \bibnamefont {Jarvis}},
		\bibinfo {author} {\bibfnamefont {P.~A.~C.}\ \bibnamefont {Brown}}, \bibinfo
		{author} {\bibfnamefont {J.~H.}\ \bibnamefont {Kim}}, \bibinfo {author}
		{\bibfnamefont {C.-H.}\ \bibnamefont {Park}}, \bibinfo {author}
		{\bibfnamefont {D.~I.}\ \bibnamefont {Khomskii}}, \bibinfo {author}
		{\bibfnamefont {S.~S.}\ \bibnamefont {Saxena}}, \ and\ \bibinfo {author}
		{\bibfnamefont {J.-G.}\ \bibnamefont {Park}},\ }\href {\doibase
		10.1103/PhysRevB.99.041402} {\bibfield  {journal} {\bibinfo  {journal} {Phys.
				Rev. B}\ }\textbf {\bibinfo {volume} {99}},\ \bibinfo {pages} {041402}
		(\bibinfo {year} {2019})}\BibitemShut {NoStop}%
	\bibitem [{\citenamefont {Lee}\ \emph {et~al.}(2016)\citenamefont {Lee},
		\citenamefont {Lee}, \citenamefont {Ryoo}, \citenamefont {Kang},
		\citenamefont {Kim}, \citenamefont {Kim}, \citenamefont {Park}, \citenamefont
		{Park},\ and\ \citenamefont {Cheong}}]{doi:10.1021/acs.nanolett.6b03052}%
	\BibitemOpen
	\bibfield  {author} {\bibinfo {author} {\bibfnamefont {J.-U.}\ \bibnamefont
			{Lee}}, \bibinfo {author} {\bibfnamefont {S.}~\bibnamefont {Lee}}, \bibinfo
		{author} {\bibfnamefont {J.~H.}\ \bibnamefont {Ryoo}}, \bibinfo {author}
		{\bibfnamefont {S.}~\bibnamefont {Kang}}, \bibinfo {author} {\bibfnamefont
			{T.~Y.}\ \bibnamefont {Kim}}, \bibinfo {author} {\bibfnamefont
			{P.}~\bibnamefont {Kim}}, \bibinfo {author} {\bibfnamefont {C.-H.}\
			\bibnamefont {Park}}, \bibinfo {author} {\bibfnamefont {J.-G.}\ \bibnamefont
			{Park}}, \ and\ \bibinfo {author} {\bibfnamefont {H.}~\bibnamefont
			{Cheong}},\ }\href {\doibase 10.1021/acs.nanolett.6b03052} {\bibfield
		{journal} {\bibinfo  {journal} {Nano Letters}\ }\textbf {\bibinfo {volume}
			{16}},\ \bibinfo {pages} {7433} (\bibinfo {year} {2016})}\BibitemShut
	{NoStop}%
	\bibitem [{\citenamefont {Huang}\ \emph {et~al.}(2017)\citenamefont {Huang},
		\citenamefont {Clark}, \citenamefont {Navarro-Moratalla}, \citenamefont
		{Klein}, \citenamefont {Cheng}, \citenamefont {Seyler}, \citenamefont
		{Zhong}, \citenamefont {Schmidgall}, \citenamefont {McGuire}, \citenamefont
		{Cobden}, \citenamefont {Yao}, \citenamefont {Xiao}, \citenamefont
		{Jarillo-Herrero},\ and\ \citenamefont {Xu}}]{ISI:000402823400033}%
	\BibitemOpen
	\bibfield  {author} {\bibinfo {author} {\bibfnamefont {B.}~\bibnamefont
			{Huang}}, \bibinfo {author} {\bibfnamefont {G.}~\bibnamefont {Clark}},
		\bibinfo {author} {\bibfnamefont {E.}~\bibnamefont {Navarro-Moratalla}},
		\bibinfo {author} {\bibfnamefont {D.~R.}\ \bibnamefont {Klein}}, \bibinfo
		{author} {\bibfnamefont {R.}~\bibnamefont {Cheng}}, \bibinfo {author}
		{\bibfnamefont {K.~L.}\ \bibnamefont {Seyler}}, \bibinfo {author}
		{\bibfnamefont {D.}~\bibnamefont {Zhong}}, \bibinfo {author} {\bibfnamefont
			{E.}~\bibnamefont {Schmidgall}}, \bibinfo {author} {\bibfnamefont {M.~A.}\
			\bibnamefont {McGuire}}, \bibinfo {author} {\bibfnamefont {D.~H.}\
			\bibnamefont {Cobden}}, \bibinfo {author} {\bibfnamefont {W.}~\bibnamefont
			{Yao}}, \bibinfo {author} {\bibfnamefont {D.}~\bibnamefont {Xiao}}, \bibinfo
		{author} {\bibfnamefont {P.}~\bibnamefont {Jarillo-Herrero}}, \ and\ \bibinfo
		{author} {\bibfnamefont {X.}~\bibnamefont {Xu}},\ }\href {\doibase
		{10.1038/nature22391}} {\bibfield  {journal} {\bibinfo  {journal} {Nature}\
		}\textbf {\bibinfo {volume} {{546}}},\ \bibinfo {pages} {{270}} (\bibinfo
		{year} {{2017}})}\BibitemShut {NoStop}%
	\bibitem [{\citenamefont {Gong}\ \emph {et~al.}(2017)\citenamefont {Gong},
		\citenamefont {Li}, \citenamefont {Li}, \citenamefont {Ji}, \citenamefont
		{Stern}, \citenamefont {Xia}, \citenamefont {Cao}, \citenamefont {Bao},
		\citenamefont {Wang}, \citenamefont {Wang}, \citenamefont {Qiu},
		\citenamefont {Cava}, \citenamefont {Louie}, \citenamefont {Xia},\ and\
		\citenamefont {Zhang}}]{ISI:000402823400032}%
	\BibitemOpen
	\bibfield  {author} {\bibinfo {author} {\bibfnamefont {C.}~\bibnamefont
			{Gong}}, \bibinfo {author} {\bibfnamefont {L.}~\bibnamefont {Li}}, \bibinfo
		{author} {\bibfnamefont {Z.}~\bibnamefont {Li}}, \bibinfo {author}
		{\bibfnamefont {H.}~\bibnamefont {Ji}}, \bibinfo {author} {\bibfnamefont
			{A.}~\bibnamefont {Stern}}, \bibinfo {author} {\bibfnamefont
			{Y.}~\bibnamefont {Xia}}, \bibinfo {author} {\bibfnamefont {T.}~\bibnamefont
			{Cao}}, \bibinfo {author} {\bibfnamefont {W.}~\bibnamefont {Bao}}, \bibinfo
		{author} {\bibfnamefont {C.}~\bibnamefont {Wang}}, \bibinfo {author}
		{\bibfnamefont {Y.}~\bibnamefont {Wang}}, \bibinfo {author} {\bibfnamefont
			{Z.~Q.}\ \bibnamefont {Qiu}}, \bibinfo {author} {\bibfnamefont {R.~J.}\
			\bibnamefont {Cava}}, \bibinfo {author} {\bibfnamefont {S.~G.}\ \bibnamefont
			{Louie}}, \bibinfo {author} {\bibfnamefont {J.}~\bibnamefont {Xia}}, \ and\
		\bibinfo {author} {\bibfnamefont {X.}~\bibnamefont {Zhang}},\ }\href
	{\doibase {10.1038/nature22060}} {\bibfield  {journal} {\bibinfo  {journal}
			{Nature}\ }\textbf {\bibinfo {volume} {{546}}},\ \bibinfo {pages} {{265}}
		(\bibinfo {year} {{2017}})}\BibitemShut {NoStop}%
	\bibitem [{\citenamefont {Joy}\ and\ \citenamefont
		{Vasudevan}(1992)}]{PhysRevB.46.5425}%
	\BibitemOpen
	\bibfield  {author} {\bibinfo {author} {\bibfnamefont {P.~A.}\ \bibnamefont
			{Joy}}\ and\ \bibinfo {author} {\bibfnamefont {S.}~\bibnamefont
			{Vasudevan}},\ }\href@noop {} {\bibfield  {journal} {\bibinfo  {journal}
			{Phys. Rev. B}\ }\textbf {\bibinfo {volume} {46}},\ \bibinfo {pages} {5425}
		(\bibinfo {year} {1992})}\BibitemShut {NoStop}%
	\bibitem [{\citenamefont {Rule}\ \emph {et~al.}(2009)\citenamefont {Rule},
		\citenamefont {Wildes}, \citenamefont {Bewley}, \citenamefont {Visser},\ and\
		\citenamefont {Hicks}}]{Rule_2009}%
	\BibitemOpen
	\bibfield  {author} {\bibinfo {author} {\bibfnamefont {K.~C.}\ \bibnamefont
			{Rule}}, \bibinfo {author} {\bibfnamefont {A.~R.}\ \bibnamefont {Wildes}},
		\bibinfo {author} {\bibfnamefont {R.~I.}\ \bibnamefont {Bewley}}, \bibinfo
		{author} {\bibfnamefont {D.}~\bibnamefont {Visser}}, \ and\ \bibinfo {author}
		{\bibfnamefont {T.~J.}\ \bibnamefont {Hicks}},\ }\href@noop {} {\bibfield
		{journal} {\bibinfo  {journal} {Journal of Physics: Condensed Matter}\
		}\textbf {\bibinfo {volume} {21}},\ \bibinfo {pages} {124214} (\bibinfo
		{year} {2009})}\BibitemShut {NoStop}%
	\bibitem [{\citenamefont {Wildes}\ \emph
		{et~al.}(2012{\natexlab{a}})\citenamefont {Wildes}, \citenamefont {Rule},
		\citenamefont {Bewley}, \citenamefont {Enderle},\ and\ \citenamefont
		{Hicks}}]{Wildes_2012}%
	\BibitemOpen
	\bibfield  {author} {\bibinfo {author} {\bibfnamefont {A.~R.}\ \bibnamefont
			{Wildes}}, \bibinfo {author} {\bibfnamefont {K.~C.}\ \bibnamefont {Rule}},
		\bibinfo {author} {\bibfnamefont {R.~I.}\ \bibnamefont {Bewley}}, \bibinfo
		{author} {\bibfnamefont {M.}~\bibnamefont {Enderle}}, \ and\ \bibinfo
		{author} {\bibfnamefont {T.~J.}\ \bibnamefont {Hicks}},\ }\href@noop {}
	{\bibfield  {journal} {\bibinfo  {journal} {Journal of Physics: Condensed
				Matter}\ }\textbf {\bibinfo {volume} {24}},\ \bibinfo {pages} {416004}
		(\bibinfo {year} {2012}{\natexlab{a}})}\BibitemShut {NoStop}%
	\bibitem [{\citenamefont {Rule}\ \emph {et~al.}(2007)\citenamefont {Rule},
		\citenamefont {McIntyre}, \citenamefont {Kennedy},\ and\ \citenamefont
		{Hicks}}]{PhysRevB.76.134402}%
	\BibitemOpen
	\bibfield  {author} {\bibinfo {author} {\bibfnamefont {K.~C.}\ \bibnamefont
			{Rule}}, \bibinfo {author} {\bibfnamefont {G.~J.}\ \bibnamefont {McIntyre}},
		\bibinfo {author} {\bibfnamefont {S.~J.}\ \bibnamefont {Kennedy}}, \ and\
		\bibinfo {author} {\bibfnamefont {T.~J.}\ \bibnamefont {Hicks}},\ }\href@noop
	{} {\bibfield  {journal} {\bibinfo  {journal} {Phys. Rev. B}\ }\textbf
		{\bibinfo {volume} {76}},\ \bibinfo {pages} {134402} (\bibinfo {year}
		{2007})}\BibitemShut {NoStop}%
	\bibitem [{\citenamefont {Wildes}\ \emph {et~al.}(2017)\citenamefont {Wildes},
		\citenamefont {Simonet}, \citenamefont {Ressouche}, \citenamefont {Ballou},\
		and\ \citenamefont {McIntyre}}]{Wildes_2017}%
	\BibitemOpen
	\bibfield  {author} {\bibinfo {author} {\bibfnamefont {A.~R.}\ \bibnamefont
			{Wildes}}, \bibinfo {author} {\bibfnamefont {V.}~\bibnamefont {Simonet}},
		\bibinfo {author} {\bibfnamefont {E.}~\bibnamefont {Ressouche}}, \bibinfo
		{author} {\bibfnamefont {R.}~\bibnamefont {Ballou}}, \ and\ \bibinfo {author}
		{\bibfnamefont {G.~J.}\ \bibnamefont {McIntyre}},\ }\href@noop {} {\bibfield
		{journal} {\bibinfo  {journal} {Journal of Physics: Condensed Matter}\
		}\textbf {\bibinfo {volume} {29}},\ \bibinfo {pages} {455801} (\bibinfo
		{year} {2017})}\BibitemShut {NoStop}%
	\bibitem [{\citenamefont {Haines}\ \emph {et~al.}(2018)\citenamefont {Haines},
		\citenamefont {Coak}, \citenamefont {Wildes}, \citenamefont {Lampronti},
		\citenamefont {Liu}, \citenamefont {Nahai-Williamson}, \citenamefont
		{Hamidov}, \citenamefont {Daisenberger},\ and\ \citenamefont
		{Saxena}}]{PhysRevLett.121.266801}%
	\BibitemOpen
	\bibfield  {author} {\bibinfo {author} {\bibfnamefont {C.~R.~S.}\
			\bibnamefont {Haines}}, \bibinfo {author} {\bibfnamefont {M.~J.}\
			\bibnamefont {Coak}}, \bibinfo {author} {\bibfnamefont {A.~R.}\ \bibnamefont
			{Wildes}}, \bibinfo {author} {\bibfnamefont {G.~I.}\ \bibnamefont
			{Lampronti}}, \bibinfo {author} {\bibfnamefont {C.}~\bibnamefont {Liu}},
		\bibinfo {author} {\bibfnamefont {P.}~\bibnamefont {Nahai-Williamson}},
		\bibinfo {author} {\bibfnamefont {H.}~\bibnamefont {Hamidov}}, \bibinfo
		{author} {\bibfnamefont {D.}~\bibnamefont {Daisenberger}}, \ and\ \bibinfo
		{author} {\bibfnamefont {S.~S.}\ \bibnamefont {Saxena}},\ }\href@noop {}
	{\bibfield  {journal} {\bibinfo  {journal} {Phys. Rev. Lett.}\ }\textbf
		{\bibinfo {volume} {121}},\ \bibinfo {pages} {266801} (\bibinfo {year}
		{2018})}\BibitemShut {NoStop}%
	\bibitem [{\citenamefont {Ressouche}\ \emph {et~al.}(2010)\citenamefont
		{Ressouche}, \citenamefont {Loire}, \citenamefont {Simonet}, \citenamefont
		{Ballou}, \citenamefont {Stunault},\ and\ \citenamefont
		{Wildes}}]{PhysRevB.82.100408}%
	\BibitemOpen
	\bibfield  {author} {\bibinfo {author} {\bibfnamefont {E.}~\bibnamefont
			{Ressouche}}, \bibinfo {author} {\bibfnamefont {M.}~\bibnamefont {Loire}},
		\bibinfo {author} {\bibfnamefont {V.}~\bibnamefont {Simonet}}, \bibinfo
		{author} {\bibfnamefont {R.}~\bibnamefont {Ballou}}, \bibinfo {author}
		{\bibfnamefont {A.}~\bibnamefont {Stunault}}, \ and\ \bibinfo {author}
		{\bibfnamefont {A.}~\bibnamefont {Wildes}},\ }\href {\doibase
		10.1103/PhysRevB.82.100408} {\bibfield  {journal} {\bibinfo  {journal} {Phys.
				Rev. B}\ }\textbf {\bibinfo {volume} {82}},\ \bibinfo {pages} {100408}
		(\bibinfo {year} {2010})}\BibitemShut {NoStop}%
	\bibitem [{\citenamefont {Grasso}\ and\ \citenamefont
		{Silipigni}(1999)}]{Grasso:99}%
	\BibitemOpen
	\bibfield  {author} {\bibinfo {author} {\bibfnamefont {V.}~\bibnamefont
			{Grasso}}\ and\ \bibinfo {author} {\bibfnamefont {L.}~\bibnamefont
			{Silipigni}},\ }\href {\doibase 10.1364/JOSAB.16.000132} {\bibfield
		{journal} {\bibinfo  {journal} {J. Opt. Soc. Am. B}\ }\textbf {\bibinfo
			{volume} {16}},\ \bibinfo {pages} {132} (\bibinfo {year} {1999})}\BibitemShut
	{NoStop}%
	\bibitem [{\citenamefont {Brec}\ \emph {et~al.}(1979)\citenamefont {Brec},
		\citenamefont {Schleich}, \citenamefont {Ouvrard}, \citenamefont {Louisy},\
		and\ \citenamefont {Rouxel}}]{Brec1979doi:10.1021/ic50197a018}%
	\BibitemOpen
	\bibfield  {author} {\bibinfo {author} {\bibfnamefont {R.}~\bibnamefont
			{Brec}}, \bibinfo {author} {\bibfnamefont {D.~M.}\ \bibnamefont {Schleich}},
		\bibinfo {author} {\bibfnamefont {G.}~\bibnamefont {Ouvrard}}, \bibinfo
		{author} {\bibfnamefont {A.}~\bibnamefont {Louisy}}, \ and\ \bibinfo {author}
		{\bibfnamefont {J.}~\bibnamefont {Rouxel}},\ }\href {\doibase
		10.1021/ic50197a018} {\bibfield  {journal} {\bibinfo  {journal} {Inorganic
				Chemistry}\ }\textbf {\bibinfo {volume} {18}},\ \bibinfo {pages} {1814}
		(\bibinfo {year} {1979})}\BibitemShut {NoStop}%
	\bibitem [{\citenamefont {Pei}\ and\ \citenamefont
		{Mi}(2019)}]{PhysRevApplied.11.014011}%
	\BibitemOpen
	\bibfield  {author} {\bibinfo {author} {\bibfnamefont {Q.}~\bibnamefont
			{Pei}}\ and\ \bibinfo {author} {\bibfnamefont {W.}~\bibnamefont {Mi}},\
	}\href {\doibase 10.1103/PhysRevApplied.11.014011} {\bibfield  {journal}
		{\bibinfo  {journal} {Phys. Rev. Applied}\ }\textbf {\bibinfo {volume}
			{11}},\ \bibinfo {pages} {014011} (\bibinfo {year} {2019})}\BibitemShut
	{NoStop}%
	\bibitem [{\citenamefont {Wiedenmann}\ \emph {et~al.}(1981)\citenamefont
		{Wiedenmann}, \citenamefont {Rossat-Mignod}, \citenamefont {Louisy},
		\citenamefont {Brec},\ and\ \citenamefont {Rouxel}}]{WIEDENMANN1981}%
	\BibitemOpen
	\bibfield  {author} {\bibinfo {author} {\bibfnamefont {A.}~\bibnamefont
			{Wiedenmann}}, \bibinfo {author} {\bibfnamefont {J.}~\bibnamefont
			{Rossat-Mignod}}, \bibinfo {author} {\bibfnamefont {A.}~\bibnamefont
			{Louisy}}, \bibinfo {author} {\bibfnamefont {R.}~\bibnamefont {Brec}}, \ and\
		\bibinfo {author} {\bibfnamefont {J.}~\bibnamefont {Rouxel}},\ }\href
	{\doibase https://doi.org/10.1016/0038-1098(81)90253-2} {\bibfield  {journal}
		{\bibinfo  {journal} {Solid State Communications}\ }\textbf {\bibinfo
			{volume} {40}},\ \bibinfo {pages} {1067 } (\bibinfo {year}
		{1981})}\BibitemShut {NoStop}%
	\bibitem [{\citenamefont {Le~Flem}\ \emph {et~al.}(1982)\citenamefont
		{Le~Flem}, \citenamefont {Brec}, \citenamefont {Ouvard}, \citenamefont
		{Louisy},\ and\ \citenamefont {Segransan}}]{ISI:A1982NN00100006}%
	\BibitemOpen
	\bibfield  {author} {\bibinfo {author} {\bibfnamefont {G.}~\bibnamefont
			{Le~Flem}}, \bibinfo {author} {\bibfnamefont {R.}~\bibnamefont {Brec}},
		\bibinfo {author} {\bibfnamefont {G.}~\bibnamefont {Ouvard}}, \bibinfo
		{author} {\bibfnamefont {A.}~\bibnamefont {Louisy}}, \ and\ \bibinfo {author}
		{\bibfnamefont {P.}~\bibnamefont {Segransan}},\ }\href {\doibase
		{10.1016/0022-3697(82)90156-1}} {\bibfield  {journal} {\bibinfo  {journal}
			{Journal of physics and chemistry of solids}\ }\textbf {\bibinfo {volume}
			{43}},\ \bibinfo {pages} {455} (\bibinfo {year} {1982})}\BibitemShut
	{NoStop}%
	\bibitem [{\citenamefont {Li}\ \emph {et~al.}(2014)\citenamefont {Li},
		\citenamefont {Wu},\ and\ \citenamefont {Yang}}]{doi:10.1021/ja505097m}%
	\BibitemOpen
	\bibfield  {author} {\bibinfo {author} {\bibfnamefont {X.}~\bibnamefont
			{Li}}, \bibinfo {author} {\bibfnamefont {X.}~\bibnamefont {Wu}}, \ and\
		\bibinfo {author} {\bibfnamefont {J.}~\bibnamefont {Yang}},\ }\href {\doibase
		10.1021/ja505097m} {\bibfield  {journal} {\bibinfo  {journal} {Journal of the
				American Chemical Society}\ }\textbf {\bibinfo {volume} {136}},\ \bibinfo
		{pages} {11065} (\bibinfo {year} {2014})}\BibitemShut {NoStop}%
	\bibitem [{\citenamefont {Liu}\ \emph {et~al.}(2020)\citenamefont {Liu},
		\citenamefont {Xu}, \citenamefont {Huang}, \citenamefont {Li}, \citenamefont
		{Feng}, \citenamefont {Huang}, \citenamefont {Zhu}, \citenamefont {Wang},
		\citenamefont {Zhang}, \citenamefont {Hou}, \citenamefont {Lu},\ and\
		\citenamefont {Xiang}}]{ISI:000522634300031}%
	\BibitemOpen
	\bibfield  {author} {\bibinfo {author} {\bibfnamefont {P.}~\bibnamefont
			{Liu}}, \bibinfo {author} {\bibfnamefont {Z.}~\bibnamefont {Xu}}, \bibinfo
		{author} {\bibfnamefont {H.}~\bibnamefont {Huang}}, \bibinfo {author}
		{\bibfnamefont {J.}~\bibnamefont {Li}}, \bibinfo {author} {\bibfnamefont
			{C.}~\bibnamefont {Feng}}, \bibinfo {author} {\bibfnamefont {M.}~\bibnamefont
			{Huang}}, \bibinfo {author} {\bibfnamefont {M.}~\bibnamefont {Zhu}}, \bibinfo
		{author} {\bibfnamefont {Z.}~\bibnamefont {Wang}}, \bibinfo {author}
		{\bibfnamefont {Z.}~\bibnamefont {Zhang}}, \bibinfo {author} {\bibfnamefont
			{D.}~\bibnamefont {Hou}}, \bibinfo {author} {\bibfnamefont {Y.}~\bibnamefont
			{Lu}}, \ and\ \bibinfo {author} {\bibfnamefont {B.}~\bibnamefont {Xiang}},\
	}\href {\doibase 10.1016/j.jallcom.2020.154432} {\bibfield  {journal}
		{\bibinfo  {journal} {Journal of Alloys and Compounds}\ }\textbf {\bibinfo
			{volume} {{828}}} (\bibinfo {year} {{2020}}),\
		10.1016/j.jallcom.2020.154432}\BibitemShut {NoStop}%
	\bibitem [{\citenamefont {Bhutani}\ \emph {et~al.}(2020)\citenamefont
		{Bhutani}, \citenamefont {Zuo}, \citenamefont {McAuliffe}, \citenamefont
		{dela Cruz},\ and\ \citenamefont {Shoemaker}}]{PhysRevMaterials.4.034411}%
	\BibitemOpen
	\bibfield  {author} {\bibinfo {author} {\bibfnamefont {A.}~\bibnamefont
			{Bhutani}}, \bibinfo {author} {\bibfnamefont {J.~L.}\ \bibnamefont {Zuo}},
		\bibinfo {author} {\bibfnamefont {R.~D.}\ \bibnamefont {McAuliffe}}, \bibinfo
		{author} {\bibfnamefont {C.~R.}\ \bibnamefont {dela Cruz}}, \ and\ \bibinfo
		{author} {\bibfnamefont {D.~P.}\ \bibnamefont {Shoemaker}},\ }\href {\doibase
		10.1103/PhysRevMaterials.4.034411} {\bibfield  {journal} {\bibinfo  {journal}
			{Phys. Rev. Materials}\ }\textbf {\bibinfo {volume} {4}},\ \bibinfo {pages}
		{034411} (\bibinfo {year} {2020})}\BibitemShut {NoStop}%
	\bibitem [{\citenamefont {Jeevanandam}\ and\ \citenamefont
		{Vasudevan}(1999)}]{Jeevanandam_1999}%
	\BibitemOpen
	\bibfield  {author} {\bibinfo {author} {\bibfnamefont {P.}~\bibnamefont
			{Jeevanandam}}\ and\ \bibinfo {author} {\bibfnamefont {S.}~\bibnamefont
			{Vasudevan}},\ }\href {\doibase 10.1088/0953-8984/11/17/314} {\bibfield
		{journal} {\bibinfo  {journal} {Journal of Physics: Condensed Matter}\
		}\textbf {\bibinfo {volume} {11}},\ \bibinfo {pages} {3563} (\bibinfo {year}
		{1999})}\BibitemShut {NoStop}%
	\bibitem [{\citenamefont {Wildes}\ \emph {et~al.}(1998)\citenamefont {Wildes},
		\citenamefont {Roessli}, \citenamefont {Lebech},\ and\ \citenamefont
		{Godfrey}}]{WildesMnPS3}%
	\BibitemOpen
	\bibfield  {author} {\bibinfo {author} {\bibfnamefont {A.~R.}\ \bibnamefont
			{Wildes}}, \bibinfo {author} {\bibfnamefont {B.}~\bibnamefont {Roessli}},
		\bibinfo {author} {\bibfnamefont {B.}~\bibnamefont {Lebech}}, \ and\ \bibinfo
		{author} {\bibfnamefont {K.~W.}\ \bibnamefont {Godfrey}},\ }\href@noop {}
	{\bibfield  {journal} {\bibinfo  {journal} {Journal of Physics: Condensed
				Matter}\ }\textbf {\bibinfo {volume} {10}},\ \bibinfo {pages} {6417}
		(\bibinfo {year} {1998})}\BibitemShut {NoStop}%
	\bibitem [{\citenamefont {Pich}\ and\ \citenamefont
		{Schwabl}(1995)}]{PICH199530}%
	\BibitemOpen
	\bibfield  {author} {\bibinfo {author} {\bibfnamefont {C.}~\bibnamefont
			{Pich}}\ and\ \bibinfo {author} {\bibfnamefont {F.}~\bibnamefont {Schwabl}},\
	}\href {\doibase https://doi.org/10.1016/0304-8853(95)00136-0} {\bibfield
		{journal} {\bibinfo  {journal} {Journal of Magnetism and Magnetic Materials}\
		}\textbf {\bibinfo {volume} {148}},\ \bibinfo {pages} {30 } (\bibinfo {year}
		{1995})}\BibitemShut {NoStop}%
	\bibitem [{\citenamefont {Hicks}\ \emph {et~al.}(2019)\citenamefont {Hicks},
		\citenamefont {Keller},\ and\ \citenamefont {Wildes}}]{HICKS2019512}%
	\BibitemOpen
	\bibfield  {author} {\bibinfo {author} {\bibfnamefont {T.}~\bibnamefont
			{Hicks}}, \bibinfo {author} {\bibfnamefont {T.}~\bibnamefont {Keller}}, \
		and\ \bibinfo {author} {\bibfnamefont {A.}~\bibnamefont {Wildes}},\ }\href
	{\doibase https://doi.org/10.1016/j.jmmm.2018.10.136} {\bibfield  {journal}
		{\bibinfo  {journal} {Journal of Magnetism and Magnetic Materials}\ }\textbf
		{\bibinfo {volume} {474}},\ \bibinfo {pages} {512 } (\bibinfo {year}
		{2019})}\BibitemShut {NoStop}%
	\bibitem [{\citenamefont {Vaclavkova{\it et al.}}(2020)}]{VaclavkovaRaman}%
	\BibitemOpen
	\bibfield  {author} {\bibinfo {author} {\bibfnamefont {D.}~\bibnamefont
			{Vaclavkova{\it et al.}}},\ }\href@noop {} {\bibfield  {journal} {\bibinfo
			{journal} {arXiv:2005.11119v1}\ } (\bibinfo {year} {2020})}\BibitemShut
	{NoStop}%
	\bibitem [{\citenamefont {Garlea}\ \emph {et~al.}(2010)\citenamefont {Garlea},
		\citenamefont {Chakoumakos}, \citenamefont {Moore}, \citenamefont {Taylor},
		\citenamefont {Chae}, \citenamefont {Maples}, \citenamefont {Riedel},
		\citenamefont {Lynn},\ and\ \citenamefont {Selby}}]{Garlea2010}%
	\BibitemOpen
	\bibfield  {author} {\bibinfo {author} {\bibfnamefont {V.~O.}\ \bibnamefont
			{Garlea}}, \bibinfo {author} {\bibfnamefont {B.~C.}\ \bibnamefont
			{Chakoumakos}}, \bibinfo {author} {\bibfnamefont {S.~A.}\ \bibnamefont
			{Moore}}, \bibinfo {author} {\bibfnamefont {G.~B.}\ \bibnamefont {Taylor}},
		\bibinfo {author} {\bibfnamefont {T.}~\bibnamefont {Chae}}, \bibinfo {author}
		{\bibfnamefont {R.~G.}\ \bibnamefont {Maples}}, \bibinfo {author}
		{\bibfnamefont {R.~A.}\ \bibnamefont {Riedel}}, \bibinfo {author}
		{\bibfnamefont {G.~W.}\ \bibnamefont {Lynn}}, \ and\ \bibinfo {author}
		{\bibfnamefont {D.~L.}\ \bibnamefont {Selby}},\ }\href {\doibase
		10.1007/s00339-010-5603-6} {\bibfield  {journal} {\bibinfo  {journal}
			{Applied Physics A}\ }\textbf {\bibinfo {volume} {99}},\ \bibinfo {pages}
		{531} (\bibinfo {year} {2010})}\BibitemShut {NoStop}%
	\bibitem [{\citenamefont {Calder}\ \emph {et~al.}(2018)\citenamefont {Calder},
		\citenamefont {An}, \citenamefont {Boehler}, \citenamefont {Dela~Cruz},
		\citenamefont {Frontzek}, \citenamefont {Guthrie}, \citenamefont {Haberl},
		\citenamefont {Huq}, \citenamefont {Kimber}, \citenamefont {Liu},
		\citenamefont {Molaison}, \citenamefont {Neuefeind}, \citenamefont {Page},
		\citenamefont {dos Santos}, \citenamefont {Taddei}, \citenamefont {Tulk},\
		and\ \citenamefont {Tucker}}]{doi:10.1063/1.5033906}%
	\BibitemOpen
	\bibfield  {author} {\bibinfo {author} {\bibfnamefont {S.}~\bibnamefont
			{Calder}}, \bibinfo {author} {\bibfnamefont {K.}~\bibnamefont {An}}, \bibinfo
		{author} {\bibfnamefont {R.}~\bibnamefont {Boehler}}, \bibinfo {author}
		{\bibfnamefont {C.~R.}\ \bibnamefont {Dela~Cruz}}, \bibinfo {author}
		{\bibfnamefont {M.~D.}\ \bibnamefont {Frontzek}}, \bibinfo {author}
		{\bibfnamefont {M.}~\bibnamefont {Guthrie}}, \bibinfo {author} {\bibfnamefont
			{B.}~\bibnamefont {Haberl}}, \bibinfo {author} {\bibfnamefont
			{A.}~\bibnamefont {Huq}}, \bibinfo {author} {\bibfnamefont {S.~A.~J.}\
			\bibnamefont {Kimber}}, \bibinfo {author} {\bibfnamefont {J.}~\bibnamefont
			{Liu}}, \bibinfo {author} {\bibfnamefont {J.~J.}\ \bibnamefont {Molaison}},
		\bibinfo {author} {\bibfnamefont {J.}~\bibnamefont {Neuefeind}}, \bibinfo
		{author} {\bibfnamefont {K.}~\bibnamefont {Page}}, \bibinfo {author}
		{\bibfnamefont {A.~M.}\ \bibnamefont {dos Santos}}, \bibinfo {author}
		{\bibfnamefont {K.~M.}\ \bibnamefont {Taddei}}, \bibinfo {author}
		{\bibfnamefont {C.}~\bibnamefont {Tulk}}, \ and\ \bibinfo {author}
		{\bibfnamefont {M.~G.}\ \bibnamefont {Tucker}},\ }\href {\doibase
		10.1063/1.5033906} {\bibfield  {journal} {\bibinfo  {journal} {Review of
				Scientific Instruments}\ }\textbf {\bibinfo {volume} {89}},\ \bibinfo {pages}
		{092701} (\bibinfo {year} {2018})}\BibitemShut {NoStop}%
	\bibitem [{\citenamefont {Arnold}\ \emph {et~al.}(2014)\citenamefont {Arnold},
		\citenamefont {Bilheux}, \citenamefont {Borreguero}, \citenamefont {Buts},
		\citenamefont {Campbell}, \citenamefont {Chapon}, \citenamefont {Doucet},
		\citenamefont {Draper}, \citenamefont {Leal}, \citenamefont {Gigg},
		\citenamefont {Lynch}, \citenamefont {Markvardsen}, \citenamefont
		{Mikkelson}, \citenamefont {Mikkelson}, \citenamefont {Miller}, \citenamefont
		{Palmen}, \citenamefont {Parker}, \citenamefont {Passos}, \citenamefont
		{Perring}, \citenamefont {Peterson}, \citenamefont {Ren}, \citenamefont
		{Reuter}, \citenamefont {Savici}, \citenamefont {Taylor}, \citenamefont
		{Taylor}, \citenamefont {Tolchenov}, \citenamefont {Zhou},\ and\
		\citenamefont {Zikovsky}}]{ARNOLD2014156}%
	\BibitemOpen
	\bibfield  {author} {\bibinfo {author} {\bibfnamefont {O.}~\bibnamefont
			{Arnold}}, \bibinfo {author} {\bibfnamefont {J.}~\bibnamefont {Bilheux}},
		\bibinfo {author} {\bibfnamefont {J.}~\bibnamefont {Borreguero}}, \bibinfo
		{author} {\bibfnamefont {A.}~\bibnamefont {Buts}}, \bibinfo {author}
		{\bibfnamefont {S.}~\bibnamefont {Campbell}}, \bibinfo {author}
		{\bibfnamefont {L.}~\bibnamefont {Chapon}}, \bibinfo {author} {\bibfnamefont
			{M.}~\bibnamefont {Doucet}}, \bibinfo {author} {\bibfnamefont
			{N.}~\bibnamefont {Draper}}, \bibinfo {author} {\bibfnamefont {R.~F.}\
			\bibnamefont {Leal}}, \bibinfo {author} {\bibfnamefont {M.}~\bibnamefont
			{Gigg}}, \bibinfo {author} {\bibfnamefont {V.}~\bibnamefont {Lynch}},
		\bibinfo {author} {\bibfnamefont {A.}~\bibnamefont {Markvardsen}}, \bibinfo
		{author} {\bibfnamefont {D.}~\bibnamefont {Mikkelson}}, \bibinfo {author}
		{\bibfnamefont {R.}~\bibnamefont {Mikkelson}}, \bibinfo {author}
		{\bibfnamefont {R.}~\bibnamefont {Miller}}, \bibinfo {author} {\bibfnamefont
			{K.}~\bibnamefont {Palmen}}, \bibinfo {author} {\bibfnamefont
			{P.}~\bibnamefont {Parker}}, \bibinfo {author} {\bibfnamefont
			{G.}~\bibnamefont {Passos}}, \bibinfo {author} {\bibfnamefont
			{T.}~\bibnamefont {Perring}}, \bibinfo {author} {\bibfnamefont
			{P.}~\bibnamefont {Peterson}}, \bibinfo {author} {\bibfnamefont
			{S.}~\bibnamefont {Ren}}, \bibinfo {author} {\bibfnamefont {M.}~\bibnamefont
			{Reuter}}, \bibinfo {author} {\bibfnamefont {A.}~\bibnamefont {Savici}},
		\bibinfo {author} {\bibfnamefont {J.}~\bibnamefont {Taylor}}, \bibinfo
		{author} {\bibfnamefont {R.}~\bibnamefont {Taylor}}, \bibinfo {author}
		{\bibfnamefont {R.}~\bibnamefont {Tolchenov}}, \bibinfo {author}
		{\bibfnamefont {W.}~\bibnamefont {Zhou}}, \ and\ \bibinfo {author}
		{\bibfnamefont {J.}~\bibnamefont {Zikovsky}},\ }\href@noop {} {\bibfield
		{journal} {\bibinfo  {journal} {Nucl. Instrum. Methods Phys. Res.}\ }\textbf
		{\bibinfo {volume} {764}},\ \bibinfo {pages} {156 } (\bibinfo {year}
		{2014})}\BibitemShut {NoStop}%
	\bibitem [{\citenamefont {Granroth}\ \emph {et~al.}(2010)\citenamefont
		{Granroth}, \citenamefont {Kolesnikov}, \citenamefont {Sherline},
		\citenamefont {Clancy}, \citenamefont {Ross}, \citenamefont {Ruff},
		\citenamefont {Gaulin},\ and\ \citenamefont
		{Nagler}}]{1742-6596-251-1-012058}%
	\BibitemOpen
	\bibfield  {author} {\bibinfo {author} {\bibfnamefont {G.~E.}\ \bibnamefont
			{Granroth}}, \bibinfo {author} {\bibfnamefont {A.~I.}\ \bibnamefont
			{Kolesnikov}}, \bibinfo {author} {\bibfnamefont {T.~E.}\ \bibnamefont
			{Sherline}}, \bibinfo {author} {\bibfnamefont {J.~P.}\ \bibnamefont
			{Clancy}}, \bibinfo {author} {\bibfnamefont {K.~A.}\ \bibnamefont {Ross}},
		\bibinfo {author} {\bibfnamefont {J.~P.~C.}\ \bibnamefont {Ruff}}, \bibinfo
		{author} {\bibfnamefont {B.~D.}\ \bibnamefont {Gaulin}}, \ and\ \bibinfo
		{author} {\bibfnamefont {S.~E.}\ \bibnamefont {Nagler}},\ }\href@noop {}
	{\bibfield  {journal} {\bibinfo  {journal} {Journal of Physics: Conference
				Series}\ }\textbf {\bibinfo {volume} {251}},\ \bibinfo {pages} {012058}
		(\bibinfo {year} {2010})}\BibitemShut {NoStop}%
	\bibitem [{\citenamefont {Azuah}\ \emph {et~al.}(2009)\citenamefont {Azuah},
		\citenamefont {Kneller}, \citenamefont {Qiu}, \citenamefont
		{Tregenna-Piggott}, \citenamefont {Brown}, \citenamefont {Copley},\ and\
		\citenamefont {Dimeo}}]{DAVE}%
	\BibitemOpen
	\bibfield  {author} {\bibinfo {author} {\bibfnamefont {R.~T.}\ \bibnamefont
			{Azuah}}, \bibinfo {author} {\bibfnamefont {L.~R.}\ \bibnamefont {Kneller}},
		\bibinfo {author} {\bibfnamefont {Y.}~\bibnamefont {Qiu}}, \bibinfo {author}
		{\bibfnamefont {P.~L.~W.}\ \bibnamefont {Tregenna-Piggott}}, \bibinfo
		{author} {\bibfnamefont {C.~M.}\ \bibnamefont {Brown}}, \bibinfo {author}
		{\bibfnamefont {J.~R.~D.}\ \bibnamefont {Copley}}, \ and\ \bibinfo {author}
		{\bibfnamefont {R.~M.}\ \bibnamefont {Dimeo}},\ }\href {\doibase
		10.6028/jres.114.025} {\bibfield  {journal} {\bibinfo  {journal} {Journal of
				Reasearch of the National Institute of Standards and Technology}\ }\textbf
		{\bibinfo {volume} {114}},\ \bibinfo {pages} {341} (\bibinfo {year}
		{2009})}\BibitemShut {NoStop}%
	\bibitem [{\citenamefont {Toth}\ and\ \citenamefont {Lake}(2015)}]{spinW}%
	\BibitemOpen
	\bibfield  {author} {\bibinfo {author} {\bibfnamefont {S.}~\bibnamefont
			{Toth}}\ and\ \bibinfo {author} {\bibfnamefont {B.}~\bibnamefont {Lake}},\
	}\href {\doibase 10.1088/0953-8984/27/16/166002} {\bibfield  {journal}
		{\bibinfo  {journal} {Journal of Physics: Condensed Matter}\ }\textbf
		{\bibinfo {volume} {27}},\ \bibinfo {pages} {166002} (\bibinfo {year}
		{2015})}\BibitemShut {NoStop}%
	\bibitem [{\citenamefont {Rodríguez-Carvajal}(1993)}]{Fullprof}%
	\BibitemOpen
	\bibfield  {author} {\bibinfo {author} {\bibfnamefont {J.}~\bibnamefont
			{Rodríguez-Carvajal}},\ }\href {\doibase
		http://dx.doi.org/10.1016/0921-4526(93)90108-I} {\bibfield  {journal}
		{\bibinfo  {journal} {Physica B: Condensed Matter}\ }\textbf {\bibinfo
			{volume} {192}},\ \bibinfo {pages} {55 } (\bibinfo {year}
		{1993})}\BibitemShut {NoStop}%
	\bibitem [{\citenamefont {Momma}\ and\ \citenamefont {Izumi}(2011)}]{VESTA}%
	\BibitemOpen
	\bibfield  {author} {\bibinfo {author} {\bibfnamefont {K.}~\bibnamefont
			{Momma}}\ and\ \bibinfo {author} {\bibfnamefont {F.}~\bibnamefont {Izumi}},\
	}\href@noop {} {\bibfield  {journal} {\bibinfo  {journal} {J. Appl.
				Crystallogr.}\ }\textbf {\bibinfo {volume} {44}},\ \bibinfo {pages} {1272}
		(\bibinfo {year} {2011})}\BibitemShut {NoStop}%
	\bibitem [{\citenamefont {Wills}(2000)}]{sarahwills}%
	\BibitemOpen
	\bibfield  {author} {\bibinfo {author} {\bibfnamefont {A.}~\bibnamefont
			{Wills}},\ }\href@noop {} {\bibfield  {journal} {\bibinfo  {journal} {Physica
				B}\ }\textbf {\bibinfo {volume} {276}},\ \bibinfo {pages} {680} (\bibinfo
		{year} {2000})}\BibitemShut {NoStop}%
	\bibitem [{\citenamefont {Perez-Mato}\ \emph {et~al.}(2015)\citenamefont
		{Perez-Mato}, \citenamefont {Gallego}, \citenamefont {Tasci}, \citenamefont
		{Elcoro}, \citenamefont {de~la Flor},\ and\ \citenamefont
		{Aroyo}}]{Bilbao_Mag}%
	\BibitemOpen
	\bibfield  {author} {\bibinfo {author} {\bibfnamefont {J.}~\bibnamefont
			{Perez-Mato}}, \bibinfo {author} {\bibfnamefont {S.}~\bibnamefont {Gallego}},
		\bibinfo {author} {\bibfnamefont {E.}~\bibnamefont {Tasci}}, \bibinfo
		{author} {\bibfnamefont {L.}~\bibnamefont {Elcoro}}, \bibinfo {author}
		{\bibfnamefont {G.}~\bibnamefont {de~la Flor}}, \ and\ \bibinfo {author}
		{\bibfnamefont {M.}~\bibnamefont {Aroyo}},\ }\href {\doibase
		10.1146/annurev-matsci-070214-021008} {\bibfield  {journal} {\bibinfo
			{journal} {Annual Review of Materials Research}\ }\textbf {\bibinfo {volume}
			{45}},\ \bibinfo {pages} {217} (\bibinfo {year} {2015})}\BibitemShut
	{NoStop}%
	\bibitem [{\citenamefont {Wildes}\ \emph {et~al.}(2015)\citenamefont {Wildes},
		\citenamefont {Simonet}, \citenamefont {Ressouche}, \citenamefont {McIntyre},
		\citenamefont {Avdeev}, \citenamefont {Suard}, \citenamefont {Kimber},
		\citenamefont {Lan\ifmmode~\mbox{\c{c}}\else \c{c}\fi{}on}, \citenamefont
		{Pepe}, \citenamefont {Moubaraki},\ and\ \citenamefont
		{Hicks}}]{PhysRevB.92.224408}%
	\BibitemOpen
	\bibfield  {author} {\bibinfo {author} {\bibfnamefont {A.~R.}\ \bibnamefont
			{Wildes}}, \bibinfo {author} {\bibfnamefont {V.}~\bibnamefont {Simonet}},
		\bibinfo {author} {\bibfnamefont {E.}~\bibnamefont {Ressouche}}, \bibinfo
		{author} {\bibfnamefont {G.~J.}\ \bibnamefont {McIntyre}}, \bibinfo {author}
		{\bibfnamefont {M.}~\bibnamefont {Avdeev}}, \bibinfo {author} {\bibfnamefont
			{E.}~\bibnamefont {Suard}}, \bibinfo {author} {\bibfnamefont {S.~A.~J.}\
			\bibnamefont {Kimber}}, \bibinfo {author} {\bibfnamefont {D.}~\bibnamefont
			{Lan\ifmmode~\mbox{\c{c}}\else \c{c}\fi{}on}}, \bibinfo {author}
		{\bibfnamefont {G.}~\bibnamefont {Pepe}}, \bibinfo {author} {\bibfnamefont
			{B.}~\bibnamefont {Moubaraki}}, \ and\ \bibinfo {author} {\bibfnamefont
			{T.~J.}\ \bibnamefont {Hicks}},\ }\href {\doibase 10.1103/PhysRevB.92.224408}
	{\bibfield  {journal} {\bibinfo  {journal} {Phys. Rev. B}\ }\textbf {\bibinfo
			{volume} {92}},\ \bibinfo {pages} {224408} (\bibinfo {year}
		{2015})}\BibitemShut {NoStop}%
	\bibitem [{\citenamefont {Wildes}\ \emph
		{et~al.}(2012{\natexlab{b}})\citenamefont {Wildes}, \citenamefont {Rule},
		\citenamefont {Bewley}, \citenamefont {Enderle},\ and\ \citenamefont
		{Hicks}}]{0953-8984-24-41-416004}%
	\BibitemOpen
	\bibfield  {author} {\bibinfo {author} {\bibfnamefont {A.~R.}\ \bibnamefont
			{Wildes}}, \bibinfo {author} {\bibfnamefont {K.~C.}\ \bibnamefont {Rule}},
		\bibinfo {author} {\bibfnamefont {R.~I.}\ \bibnamefont {Bewley}}, \bibinfo
		{author} {\bibfnamefont {M.}~\bibnamefont {Enderle}}, \ and\ \bibinfo
		{author} {\bibfnamefont {T.~J.}\ \bibnamefont {Hicks}},\ }\href
	{http://stacks.iop.org/0953-8984/24/i=41/a=416004} {\bibfield  {journal}
		{\bibinfo  {journal} {Journal of Physics: Condensed Matter}\ }\textbf
		{\bibinfo {volume} {24}},\ \bibinfo {pages} {416004} (\bibinfo {year}
		{2012}{\natexlab{b}})}\BibitemShut {NoStop}%
	\bibitem [{\citenamefont {Lan\ifmmode~\mbox{\c{c}}\else \c{c}\fi{}on}\ \emph
		{et~al.}(2018)\citenamefont {Lan\ifmmode~\mbox{\c{c}}\else \c{c}\fi{}on},
		\citenamefont {Ewings}, \citenamefont {Guidi}, \citenamefont {Formisano},\
		and\ \citenamefont {Wildes}}]{PhysRevB.98.134414}%
	\BibitemOpen
	\bibfield  {author} {\bibinfo {author} {\bibfnamefont {D.}~\bibnamefont
			{Lan\ifmmode~\mbox{\c{c}}\else \c{c}\fi{}on}}, \bibinfo {author}
		{\bibfnamefont {R.~A.}\ \bibnamefont {Ewings}}, \bibinfo {author}
		{\bibfnamefont {T.}~\bibnamefont {Guidi}}, \bibinfo {author} {\bibfnamefont
			{F.}~\bibnamefont {Formisano}}, \ and\ \bibinfo {author} {\bibfnamefont
			{A.~R.}\ \bibnamefont {Wildes}},\ }\href@noop {} {\bibfield  {journal}
		{\bibinfo  {journal} {Phys. Rev. B}\ }\textbf {\bibinfo {volume} {98}},\
		\bibinfo {pages} {134414} (\bibinfo {year} {2018})}\BibitemShut {NoStop}%
	\bibitem [{\citenamefont {Kim}\ \emph {et~al.}(2020)\citenamefont {Kim},
		\citenamefont {Jeong}, \citenamefont {Masuda}, \citenamefont {Asai},
		\citenamefont {Itoh}, \citenamefont {Kim}, \citenamefont {Wildes},\ and\
		\citenamefont {Park}}]{INSCoPS3}%
	\BibitemOpen
	\bibfield  {author} {\bibinfo {author} {\bibfnamefont {C.}~\bibnamefont
			{Kim}}, \bibinfo {author} {\bibfnamefont {J.}~\bibnamefont {Jeong}}, \bibinfo
		{author} {\bibfnamefont {T.}~\bibnamefont {Masuda}}, \bibinfo {author}
		{\bibfnamefont {S.}~\bibnamefont {Asai}}, \bibinfo {author} {\bibfnamefont
			{S.}~\bibnamefont {Itoh}}, \bibinfo {author} {\bibfnamefont {H.-S.}\
			\bibnamefont {Kim}}, \bibinfo {author} {\bibfnamefont {A.}~\bibnamefont
			{Wildes}}, \ and\ \bibinfo {author} {\bibfnamefont {J.-G.}\ \bibnamefont
			{Park}},\ }\href@noop {} {\bibfield  {journal} {\bibinfo  {journal}
			{arXiv:2007.11791}\ } (\bibinfo {year} {2020})}\BibitemShut {NoStop}%
	\bibitem [{\citenamefont {Sivadas}\ \emph {et~al.}(2015)\citenamefont
		{Sivadas}, \citenamefont {Daniels}, \citenamefont {Swendsen}, \citenamefont
		{Okamoto},\ and\ \citenamefont {Xiao}}]{PhysRevB.91.235425}%
	\BibitemOpen
	\bibfield  {author} {\bibinfo {author} {\bibfnamefont {N.}~\bibnamefont
			{Sivadas}}, \bibinfo {author} {\bibfnamefont {M.~W.}\ \bibnamefont
			{Daniels}}, \bibinfo {author} {\bibfnamefont {R.~H.}\ \bibnamefont
			{Swendsen}}, \bibinfo {author} {\bibfnamefont {S.}~\bibnamefont {Okamoto}}, \
		and\ \bibinfo {author} {\bibfnamefont {D.}~\bibnamefont {Xiao}},\ }\href
	{\doibase 10.1103/PhysRevB.91.235425} {\bibfield  {journal} {\bibinfo
			{journal} {Phys. Rev. B}\ }\textbf {\bibinfo {volume} {91}},\ \bibinfo
		{pages} {235425} (\bibinfo {year} {2015})}\BibitemShut {NoStop}%
	\bibitem [{\citenamefont {Yang}\ \emph {et~al.}(2020)\citenamefont {Yang},
		\citenamefont {Zhou}, \citenamefont {Guo}, \citenamefont {Dedkov},\ and\
		\citenamefont {Voloshina}}]{C9RA09030D}%
	\BibitemOpen
	\bibfield  {author} {\bibinfo {author} {\bibfnamefont {J.}~\bibnamefont
			{Yang}}, \bibinfo {author} {\bibfnamefont {Y.}~\bibnamefont {Zhou}}, \bibinfo
		{author} {\bibfnamefont {Q.}~\bibnamefont {Guo}}, \bibinfo {author}
		{\bibfnamefont {Y.}~\bibnamefont {Dedkov}}, \ and\ \bibinfo {author}
		{\bibfnamefont {E.}~\bibnamefont {Voloshina}},\ }\href {\doibase
		10.1039/C9RA09030D} {\bibfield  {journal} {\bibinfo  {journal} {RSC Adv.}\
		}\textbf {\bibinfo {volume} {10}},\ \bibinfo {pages} {851} (\bibinfo {year}
		{2020})}\BibitemShut {NoStop}%
	\bibitem [{\citenamefont {Chittari}\ \emph {et~al.}(2016)\citenamefont
		{Chittari}, \citenamefont {Park}, \citenamefont {Lee}, \citenamefont {Han},
		\citenamefont {MacDonald}, \citenamefont {Hwang},\ and\ \citenamefont
		{Jung}}]{PhysRevB.94.184428}%
	\BibitemOpen
	\bibfield  {author} {\bibinfo {author} {\bibfnamefont {B.~L.}\ \bibnamefont
			{Chittari}}, \bibinfo {author} {\bibfnamefont {Y.}~\bibnamefont {Park}},
		\bibinfo {author} {\bibfnamefont {D.}~\bibnamefont {Lee}}, \bibinfo {author}
		{\bibfnamefont {M.}~\bibnamefont {Han}}, \bibinfo {author} {\bibfnamefont
			{A.~H.}\ \bibnamefont {MacDonald}}, \bibinfo {author} {\bibfnamefont
			{E.}~\bibnamefont {Hwang}}, \ and\ \bibinfo {author} {\bibfnamefont
			{J.}~\bibnamefont {Jung}},\ }\href {\doibase 10.1103/PhysRevB.94.184428}
	{\bibfield  {journal} {\bibinfo  {journal} {Phys. Rev. B}\ }\textbf {\bibinfo
			{volume} {94}},\ \bibinfo {pages} {184428} (\bibinfo {year}
		{2016})}\BibitemShut {NoStop}%
	\bibitem [{\citenamefont {Pei}\ \emph {et~al.}(2018)\citenamefont {Pei},
		\citenamefont {Wang}, \citenamefont {Zou},\ and\ \citenamefont
		{Mi}}]{ISI:000433428500001}%
	\BibitemOpen
	\bibfield  {author} {\bibinfo {author} {\bibfnamefont {Q.}~\bibnamefont
			{Pei}}, \bibinfo {author} {\bibfnamefont {X.-C.}\ \bibnamefont {Wang}},
		\bibinfo {author} {\bibfnamefont {J.-J.}\ \bibnamefont {Zou}}, \ and\
		\bibinfo {author} {\bibfnamefont {W.-B.}\ \bibnamefont {Mi}},\ }\href
	{\doibase {10.1007/s11467-018-0796-9}} {\bibfield  {journal} {\bibinfo
			{journal} {Frontiers of Physics}\ }\textbf {\bibinfo {volume} {{13}}}
		(\bibinfo {year} {{2018}}),\ {10.1007/s11467-018-0796-9}}\BibitemShut
	{NoStop}%
\end{thebibliography}

%

\end{document}